\documentclass[a4paper,11pt]{article}
\pdfoutput=1
\usepackage{heppub}
\usepackage{amsthm}
\usepackage{amsmath}
\usepackage{bbm}
\usepackage{graphicx,accents}
\usepackage{slashed}
\usepackage{amssymb}
\usepackage{mathrsfs}
\usepackage{footmisc}
\usepackage{enumitem}
\usepackage[usenames,dvipsnames]{xcolor}
\usepackage{braket}
\usepackage{bm,subcaption}
\usepackage{xfrac}
\usepackage{comment}

\newcommand{\pzp}{\nu}

\usepackage{marginnote}
\usepackage[normalem]{ulem}

\author[1,2]{Jean F. Du Plessis,}
\author[2]{Bruno Scheihing-Hitschfeld}
\author[1,2]{and Rachel Steinhorst}

\affiliation[1]{MIT Center for Theoretical Physics -- a Leinweber Institute, Massachusetts Institute of Technology, Cambridge, MA  02139, USA}

\affiliation[2]{Kavli Institute for Theoretical Physics, University of California Santa Barbara, Santa Barbara, California 93106, USA}

\title{Attractodynamics in \(0+1\)D}

\preprint{MIT-CTP/6068}

\abstract{
Hydrodynamics is a macroscopic theory of long-wavelength dynamics around local thermal equilibrium.  We develop \textit{attractodynamics}, the analogous construction around a far-from-equilibrium attractor. 
Dynamics near a far-from-equilibrium attractor retains some non-hydrodynamic microscopic information, which attractodynamics systematically organizes.
We move towards this general structure by starting in $0+1$D, and consider a model for which an anisotropic far-from-equilibrium attractor solution is exactly known. This setting provides a clean benchmark in which the ideal attractodynamic equations and a leading transient residual extension can be compared directly with the full kinetic evolution.  The resulting hierarchy gives an improvable description of near-attractor dynamics: the ideal theory captures evolution close to the attracting manifold, while retaining the leading off-attractor moments extends the regime of agreement until the finite truncation breaks down. This example identifies the ingredients needed for local \(3+1\)D attractodynamics and for macroscopic attractodynamic theories not derived from an underlying kinetic description.
}

\emailAdd{jeandp@mit.edu}
\emailAdd{bscheihi@kitp.ucsb.edu}
\emailAdd{rstein99@mit.edu}

\begin{document}

\maketitle

\section{Introduction}
\label{sec:Intro}

Understanding the dynamics of quantum many-body systems far from equilibrium
is a central problem in many areas of physics, ranging from the quark-gluon
plasma produced in heavy-ion collisions \cite{Heinz:2001xi,Heinz:2002un,Kolb:2003dz,Romatschke:2003ms,Heinz:2004pj,Chesler:2008hg,Chesler:2009cy,Chesler:2010bi,Heller:2011ju,Heller:2012je,Heller:2012km,vanderSchee:2012qj,Schlichting:2012es,Kurkela:2012hp,Berges:2013eia,Berges:2013fga,Berges:2013lsa,AbraaoYork:2014hbk,Kurkela:2015qoa,Heller:2016rtz,Florkowski:2016zsi,Romatschke:2017vte,Alqahtani:2017mhy,Kurkela:2018wud,Schlichting:2019abc,Berges:2020fwq,Mitra:2020mei} to ultracold atomic gases~\cite{PineiroOrioli:2015cpb,Mikheev:2018adp,Prufer:2018hto,Erne:2018gmz,Glidden:2020qmu,Huh:2023xso,Gazo:2023exc,Lannig:2023fzf,Martirosyan:2023mml} 
non-equilibrium quantum field theory more broadly \cite{Micha:2002ey,Berges:2004yj,Berges:2015kfa}, and cosmological defect
networks such as axion strings
\cite{Gorghetto:2018myk,Hindmarsh:2021vih}.
Although the microscopic descriptions of these systems may be very different,
their late-time, long-wavelength dynamics is often much simpler.  When no
additional parametrically slow modes are present, the late-time dynamics near
thermal equilibrium is described by hydrodynamics.

Hydrodynamics is powerful because it is a universal macroscopic theory organized around
local thermal equilibrium.  It does not require a quasiparticle description or
a particular form of their interactions; microscopic information enters through an
equation of state and transport coefficients.  In this sense hydrodynamics is not merely the statement that
thermal equilibrium is a late-time attractor.  It is the local effective
theory of long-wavelength dynamics around that attractor
\cite{Kovtun:2012rj,Crossley:2015evo,Liu:2018kfw}.

It is therefore natural to look for earlier regimes in which much of the
memory of the initial condition has already been lost, but while the system is
still far from local thermal equilibrium.  Such regimes are commonly described
in terms of attractor behavior: a broad class of initial conditions is driven
toward the same lower-dimensional region in the space of possible states on a
timescale shorter than the full thermal equilibration time.  This idea has
played an important role in understanding the rapid onset of hydrodynamic
behavior in heavy-ion collisions, both in weakly coupled kinetic descriptions~\cite{Mueller:1999fp,Mueller:1999pi,Mueller:2002gd,Baier:2000sb,Arnold:2002zm,Arnold:2003rq,Kurkela:2014tea,Kurkela:2015qoa,Kurkela:2018vqr,Kurkela:2018wud,Blaizot:2011xf,Blaizot:2013lga,Tanji:2017suk,Mazeliauskas:2018yef,Boguslavski:2023jvg} 
and in strongly coupled holographic models~\cite{Chesler:2008hg,Chesler:2009cy,Chesler:2010bi,Heller:2012je,Heller:2012km,vanderSchee:2012qj,Heller:2013oxa,Casalderrey-Solana:2011dxg,Chesler:2015lsa,Chesler:2015fpa,Heller:2016gbp,Chesler:2015bba,Chesler:2016ceu,Grozdanov:2016zjj,Folkestad:2019lam,Mitra:2020mei}.
In this context, attractors provide a useful language for separating the loss
of memory of generic initial conditions from the later approach to local
thermal equilibrium.

Recent works have argued that fluid dynamics is applicable near an isotropic attractor \cite{Denicol:2021wod,Berges:2025ccd} using methods similar to the ones we will apply here. Here we ask a more general question:
 whether generic anisotropic far-from-equilibrium
attractors can be used not merely as distinguished phase space trajectories or loci
selected by a microscopic dynamical system, but as organizing structures for a 
macroscopic theory of long-wavelength dynamics.  
A macroscopic theory should specify the local
variables, the equations that evolve them, the constitutive data needed to
close those equations, and the residual modes that describe controlled
departures from the leading family of states. In this work, we ask whether this structure
can be built around a far-from-equilibrium attractor rather than around local
thermal equilibrium, and give an affirmative answer with a specific example.

Closely related ideas appear in anisotropic hydrodynamics and in moment-based
descriptions of far-from-equilibrium kinetic theory
\cite{Israel:1979wp,Alqahtani:2017mhy,Florkowski:2016zsi,
Romatschke:2003ms,Romatschke:2016hle,Romatschke:2017vte}.  These frameworks
promote parameters of a non-equilibrium distribution to dynamical fields and
can give accurate descriptions of systems with large momentum-space
anisotropies.  In simple symmetry reductions, anisotropic hydrodynamics also
has attractor solutions which closely track the attractors of kinetic theory.
Thus the novelty of the present work is not the use of anisotropic
one-particle distributions, nor the existence of attractor solutions in
hydrodynamic or kinetic models.  The distinction is one of organization: we
take the attracting manifold itself to be the object around which the
macroscopic theory is built.

We call this organization \emph{attractodynamics}.  In the kinetic-theory
realization developed here, an attractor manifold is defined as a sub-space in the space of distribution functions, this is furnished with a coordinate chart and is supplemented by matching
conditions, which separate motion along the attracting manifold from residual off-attractor perturbations.  The leading, or ideal, theory evolves only the chart variables which parametrize the attractor surface.  Systematic improvements may either eliminate residual perturbations in a gradient expansion or retain selected residual moments as transient degrees of freedom, in analogy with Israel--Stewart-type extensions of hydrodynamics. Such a truncation is useful only when the omitted off-attractor modes relax rapidly, or are weakly sourced, on the timescales over which the chart
variables evolve.

The goal of this paper is twofold.  First, we give a controlled proof of
principle in a symmetry-reduced kinetic setting.  We use the $0+1$D diffusion model studied by
Brewer, Scheihing-Hitschfeld and Yin (BSY)~\cite{Brewer:2022vkq}, a boost-invariant kinetic
equation with a known exact nonthermal attractor, even away from the so called BMSS (scaling) fixed point~\cite{Baier:2000sb}. Furthermore, this fixed point has also been observed in classical gauge theory simulations~\cite{Berges:2013eia,Berges:2013fga}. Owing to the exact results for the attractor found by BSY, we shall refer to it as the BSY model.  This model is simple enough that the exact attractor
chart, tangent directions, matching conditions, and leading residual sector
can be written explicitly, while still allowing a direct comparison between
the reduced attractodynamic evolution and the full kinetic equation. For more general theories, characterizing the far-from-equilibrium attractor may be possible using the Adiabatic Hydrodynamization framework~\cite{Brewer:2019oha,Rajagopal:2024lou,Rajagopal:2025nca}. Adiabatic Hydrodynamization can be used to understand the approach to the attractor and its structure, whereas attractodynamics is a tool to directly study non-thermal flow-like behavior in comparison to hydrodynamics. In the language of effective field theory, Adiabatic Hydrodynamization provides a top-down approach to identifying the attractor, whereas attractodynamics only describes the low-energy dynamics around the attractor, and could in principle be constructed in a bottom-up fashion. That being said, note that because we restrict to $0+1$D, the examination of sound modes is beyond the scope of this paper. Second, we use this construction to identify the macroscopic data underlying attractodynamics.  In a kinetic realization these data are obtained from moments of the distribution function, matching conditions, and projected kinetic equations.  In the macroscopic formulation they are the variables and constitutive data of the theory itself: chart variables, fluxes, source terms, and residual relaxation data.  The local spacetime-dependent theory is then obtained by promoting the symmetry-reduced scalar data to tensorial decompositions and gradient structures.

The paper is organized as follows.  In
Sec.~\ref{sec:general-attractodynamics} we discuss how to formulate ideal and viscous attractodynamics starting from a kinetic theory setting, using the attractor manifold as the fundamental input and imposing appropriate matching conditions to determine the dynamics, and separate the viscous and ideal degrees of freedom.
We then implement this construction concretely in the BSY
model in Sec.~\ref{sec:BSY}, derive the \(0+1\)D ideal and viscous equations, and compare their evolution to each other and to that of the full kinetic theory evolution. Finally, in Sec.~\ref{sec:attract-beyond} we discuss how attractodynamics may be detached from the microscopic physics that gives rise to it. That is to say, we discuss how such a theory may be constructed from a macroscopic point of view, without assuming an underlying kinetic theory description. 
We present our conclusions and discuss future directions in Sec.~\ref{sec:conclusions}.

\paragraph{Notation}

Throughout this paper, we will use \begin{align}
    \braket{\psi}_f \equiv  \int \frac{d^3p}{(2\pi)^3 E} \psi f
\end{align}
where $E=p\cdot u$. We will also use the spacelike projector
\begin{equation}
    \Delta^{\mu\nu} \equiv g^{\mu\nu}-u^\mu u^\nu
\end{equation}
and the transverse projector
\begin{equation}
    \Xi^{\mu\nu} \equiv \Delta^{\mu\nu} + \ell^\mu \ell^\nu
\end{equation}
where $\ell$ refers to some preferred direction of anisotropy (such that $\ell^2 = -1$ and $\ell \cdot u = 0$). We will also find it convenient to write the spacelike traceless part $A^{\langle\mu\nu \rangle}$
of a tensor $A$ as 
\begin{equation}
    A^{\langle \mu \nu\rangle}\equiv\Delta^{\mu\nu}_{\alpha\beta}A^{\alpha\beta} 
    \end{equation}
    where 
\begin{equation}\Delta^{\mu\nu}_{\alpha\beta}\equiv\Delta^{(\mu}_\alpha\Delta^{\nu)}_\beta-\frac13\Delta^{\mu\nu}\Delta_{\alpha\beta}.
\end{equation}

\section{Attractors and Attractodynamics}
\label{sec:general-attractodynamics}

The stated goal of our paper requires us to consider physical systems described by equations of motion that exhibit ``attractor'' solutions. We will focus on kinetic theories, which are flexible enough to describe a broad class of many-body phenomena, including attractor behavior, the process of hydrodynamization, and hydrodynamics. They define a tractable class of microscopic theories with non-trivial out-of-equilibrium dynamics, and as such may be used as a testing bed in the development of out-of-equilibrium EFTs. 

A kinetic theory is specified by a distribution function $f(x,p)$ describing the density of particles at spacetime position $x = (t,{\bf x})$ with 4-momentum $p = (p^0,{\bf p})$, where $p^0$ is determined via a dispersion relation $p^0 = p^0({\bf p})$, and a kinetic equation
\begin{equation}
    p^\mu \frac{\partial}{\partial x^\mu} f(x,p) = - \mathcal{C}[f](x,p) \, , \label{eq:kin}
\end{equation}
characterized by a collision kernel $\mathcal{C}[f]$, which in principle is an arbitrary functional of $f$, up to conservation equations imposed by symmetries of the microscopic interactions.\footnote{In coordinates with a nontrivial connection the partial derivative gets promoted to a covariant derivative.}

In this context, we will \textit{define} an attractor as a distribution function $f(x,p)$ that solves the kinetic equation~\eqref{eq:kin}, and additionally satisfies 2 requirements:
\begin{enumerate}
    \item It is characterized by a functional form $f_{\mathcal{A}}(p,\phi^I)$ and $N$ spacetime-dependent parameters $\phi_{\mathcal{A}}^I(x)$ ($I \in \{1,\ldots,N\}$) satisfying a set of \textit{attractor evolution equations}
    \begin{equation}
        \mathcal{F}_{\mathcal{A}}^J( \partial_\mu \phi, \phi,x ) = 0 \, , \label{eq:attractor-manifold-evol}
    \end{equation}
    where $\mathcal{F}_{\mathcal{A}}^J$ are functions (not functionals) of $\phi^I$ and their first derivatives,
    such that
    \begin{equation}
        f(x,p) = f_{\mathcal{A}}(p;\phi_{\mathcal{A}}^I(x))
    \end{equation}
    solves the kinetic equation for \textit{any} solution $\phi_{\mathcal{A}}^I(x)$ to the evolution equations~\eqref{eq:attractor-manifold-evol}. There can be more than $N$ equations $\mathcal{F}^J = 0$, in which case the surplus above $N$ should be interpreted as constraints\footnote{For example, isotropy for a static thermal distribution as an attractor. }. 
    \item The state is stable under perturbations orthogonal to the tangent space of the attractor manifold $\{\partial f_{\mathcal{A}}/\partial \phi^I\}_I$. A precise definition of this condition requires the introduction of projection operators; it is convenient to choose them once the collision kernel and boundary conditions are explicit. 
\end{enumerate}
An attractor therefore contains two distinct pieces of structure.  First,
the parametrized family \(f_{\mathcal A}(p;\phi^I)\) defines a
manifold of distributions, which we call the \emph{attractor manifold}, described by
coordinates \(\phi^I\).  Second, the attractor evolution equations
\(\mathcal F_{\mathcal A}^J=0\) select the spacetime-dependent sections
\(\phi_{\mathcal A}^I(x)\) for which
\(f_{\mathcal A}(p;\phi_{\mathcal A}^I(x))\) is an exact solution of the
kinetic equation.  

A general choice of fields \(\phi^I(x)\) is therefore
pointwise on the attractor chart, but only defines an exact attractor solution 
when it satisfies Eq.~\eqref{eq:attractor-manifold-evol}. However, one will in general have to deal with approximate attractors. 
In these cases, a systematic expansion may be constructed by developing Taylor series of $\ln f_\mathcal{A}$ about its maximum, keeping terms up to the desired accuracy.
The simplest attractor is specified by the thermal distribution
\begin{equation}
    \ln f_{\mathcal{A},\rm{therm}}(p^\mu;\alpha,\beta,u^\mu)\equiv\alpha-\beta p\cdot u,
\end{equation}
which is the familiar late-time thermal attractor in kinetic theories that relax to equilibrium. 
Since this is a static and isotropic attractor, the attractor evolution equations are given by
\begin{equation}
    \partial_\mu \alpha=\partial_\mu\beta=\partial_\mu u^\nu=0.
\end{equation}
The dynamical evolution of $\alpha,\beta,u^\mu$ when spatial gradients are introduced (the spatially dependent evolution on the attractor manifold) is described, at leading order in the hydrodynamic gradient expansion, by ideal hydrodynamics; whose analog we will explore below. In this way, a system being locally on the attractor manifold is directly analogous to the system being in local thermal equilibrium.

\subsection{Local Dynamics on the Attractor Manifold}

Consider now the case where the distribution function is close to, but not exactly, locally on the attractor manifold. More precisely, assume that the distribution function is of the form
\begin{equation}
    f(x,p) = f_{\mathcal A}(p;\phi^I(x) ) + \delta f(x,p) \, , \label{eq:fA-perturbed}
\end{equation}
where $\phi^I(x)$ is not necessarily an element of the solutions to the attractor evolution equations $\phi_{\mathcal A}^I$. 
Equation~\eqref{eq:fA-perturbed} is the analog of considering perturbations on top of a thermal distribution with a spacetime-dependent velocity and temperature fields, which yields hydrodynamics.

\paragraph{Ideal Attractodynamics}

The analog of ideal hydrodynamics is obtained when we consider a theory of the variables $\phi^I$ describing a state that is locally on the attractor. 
Operationally, this means to consider $\delta f \to 0$ in Eq.~\eqref{eq:fA-perturbed}, and \textit{assume} it remains negligible in the dynamics.

We expect that typically there exists a non-unique set of moments naturally associated with the chart variables $\phi^I$ of the form
\begin{equation} \label{eq:associated-moments}
    \Big\langle p\cdot u \frac{\partial\ln f_\mathcal{A}}{\partial\phi^I}\Big\rangle_f.
\end{equation}
These represent moments of functions which span the tangent space of the attractor manifold, and furthermore, whenever $\phi^I$ are coefficients in a polynomial expansion of $\ln f_\mathcal{A}$ these are then understood as components of
\begin{equation}
    I^{\lambda\mu_1\cdots\mu_n}[f]
    =
    \int \frac{d^3p}{(2\pi)^3(u\cdot p)}
    p^\lambda p^{\mu_1}\cdots p^{\mu_n} f \,.
    \label{eq:local-moment-hierarchy}
\end{equation}
For example, in hydrodynamics, $\mu$ and $T$ are associated with the macroscopic densities
\begin{align}
    n(\alpha,\beta) &= u_\mu N^\mu\,,
    \label{eq:local-number-density}\\
    \mathcal E(\alpha,\beta) &= u_\mu u_\nu T^{\mu\nu}\,,
    \label{eq:local-energy-density}
\end{align}
where we can note that $N^\mu=I^\mu$ and $T^{\mu\nu}=I^{\mu\nu}$ in the notation of Eq.~\eqref{eq:local-moment-hierarchy}. We therefore find it useful to decompose the tensors $N^\mu,T^{\mu\nu}$ in terms of $u^\mu$ and $\Delta^{\mu\nu}$. In the Landau frame,
\begin{align}
    N^\mu &= \braket{E} u^\mu + \braket{p_\nu}\Delta^{\mu\nu}\\
    T^{\mu\nu} &= \braket{E^2} u^\mu u^\nu + \braket{p^\alpha p^\beta} \Delta^{\mu\nu}_{\alpha\beta} + \frac{1}{3} \braket{p^\alpha p^\beta \Delta_{\alpha \beta}}\Delta^{\mu\nu} \,.
\end{align}\
With an anisotropic system, it may instead be useful to decompose in terms of $u^\mu$, $\ell^\mu$, and $\Xi^{\mu\nu}$, in which case we have
\begin{align}
    N^\mu &= \braket{E} u^\mu + \braket{p_\parallel} \ell^\mu + \braket{p_\nu}\Xi^{\mu\nu} \\
    T^{\mu\nu} &= \braket{E^2} u^\mu u^\nu + \braket{p_\alpha p_\beta} \Xi^{\mu\alpha}\Xi^{\nu\beta} + \braket{p_\parallel p_\parallel } \ell^\mu \ell^\nu + 2 \braket{p_\parallel p_\alpha} \ell^{(\mu} \Xi^{\nu) \alpha}\,,
\end{align}
where we have defined $p_\parallel=-\ell_\mu p^\mu$. The final term in $T^{\mu\nu}$ can be omitted if the system is even in the longitudinal direction $\ell^\mu$, and the final term of $N^\mu$ can be omitted if the system is rotationally symmetric in the transverse plane.
In attractodynamics we will generally also consider moments higher than $T^{\mu\nu}$ to capture the non-hydrodynamic degrees of freedom. For example, (assuming the system is even in $
\ell^\mu$ for simplicity) we can decompose $I^{\lambda \mu\nu}$ as
\begin{align}
\begin{split}
    I^{\lambda\mu\nu} &= \braket{E^3}u^\lambda u^\mu u^\nu + \braket{E p_\alpha p_\beta} (u^\lambda \Xi^{\alpha \mu} \Xi^{\beta \nu} + u^\nu \Xi^{\alpha \lambda} \Xi^{\beta \mu} + u^\mu \Xi^{\alpha \lambda} \Xi^{\beta \nu} ) \\
    &\qquad + \braket{E p_\parallel^2} (u^\lambda \ell^\mu \ell^\nu + u^\mu \ell^\lambda \ell^\nu + u^\nu \ell^\lambda \ell^\mu) + \braket{p_\alpha p_\beta p_\gamma} (\Xi^{\alpha \lambda} \Xi^{\beta \mu} \Xi^{\gamma \nu}) \\
    &\qquad + \braket{p_\alpha p_\parallel^2 } (\Xi^{\alpha \lambda} \ell^\mu \ell^\nu + \Xi^{\alpha \mu} \ell^\lambda \ell^\nu + \Xi^{\alpha \nu} \ell^\lambda \ell^\mu) \\
    &\qquad + \braket{E^2 p_\alpha}(u^\lambda u^\mu \Xi^{\alpha \nu} + u^\mu u^\nu \Xi^{\alpha \lambda} + u^\lambda u^\nu \Xi^{\alpha \mu}  ) \,. \label{eq:ilambdamunu-decomposition}
\end{split}
\end{align}
Additionally, assuming rotational symmetry in the transverse plane on the attractor, the final three terms of $I^{\lambda\mu\nu}$ must vanish.
The dynamics of the attractor variables $\phi^I$ can then be determined by taking moments of the Boltzmann equation,
\begin{equation} 
    \partial_\lambda I^{\lambda\mu_1\cdots\mu_n}
    =-
    \mathcal C^{\mu_1\cdots\mu_n}\,,
    \label{eq:local-moment-eom}
\end{equation}
where \(\mathcal C^{\mu_1\cdots\mu_n}\) denotes the corresponding moment of
the collision kernel, 
\begin{equation}
    \mathcal C^{\mu_1\cdots\mu_n}[f]\equiv
\int \frac{d^3p}{(2\pi)^3(u\cdot p)}
p^{\mu_1}\cdots p^{\mu_n}\mathcal C[f]\,.
\end{equation}

Note that Eq.~\eqref{eq:local-moment-eom} implicitly contains \textit{generalized equations of state} for the macroscopic variables in the currents, as the attractor $f_{\mathcal A}$ determines all components of the current, not just the density being matched and evolved.
Whether Eq.~\eqref{eq:local-moment-eom} is a good approximation of the evolution of the moments in the underlying kinetic theory depends on how large the correction is to all of its terms due to a small non-vanishing $\delta f$, which depends on the details of the underlying kinetic theory\footnote{Specifically, whether matching conditions can be chosen such that the effect of perturbations $\delta f$ on the chosen moments remain small compared to the dynamics of $\phi^I$ if we solved the full kinetic theory dynamics --- as they should provided the distribution function is close to the attractor.}.

\paragraph{Matching conditions}

Given that, as in the hydrodynamic case, the attractor component does not in general solve the kinetic equation by itself, it is desirable to give physical meaning to the attractor component in Eq.~\eqref{eq:fA-perturbed}. It is convenient to choose the splitting between $f_\mathcal{A}$ and $\delta f$ in this equation by requiring that $f_{\mathcal A}$ carries all of the information required to determine a chosen set of (local) macroscopic variables, typically of the form $u_\lambda R_{\mu_1\cdots\mu_n}  I^{\lambda\mu_1\cdots\mu_n}$ where $R$ is a tensor composed of $u^\mu$, $\ell^\mu$, $\Xi^{\mu\nu}$. In hydrodynamics this is equivalent to specifying a choice of frame.

In general, one will consider $N$ matching conditions, specified by functionals $\{M_J\}_{J=1}^N$ of the distribution function $f$, which impose $N$ conditions on $\delta f$ as 
\begin{equation}
    M_J[f_{\mathcal{A}} + \delta f ] = M_J[f_{\mathcal{A}}] \, , \quad J \in \{1,\ldots,N\} \, .
\end{equation}
These matching conditions fix the attractor variables $\phi^I$, as do the Landau conditions for $T$ and $u$ in hydrodynamics. In practice, we will mainly consider matching conditions $M_J$ that are linear functionals of $f$, and therefore the requirements on $\delta f$ become
\begin{equation}
    M_J^{\rm linear}[\delta f] = 0 \, .
\end{equation}
The number of matching conditions needed is equal to the number of dynamical variables $\phi^I$ in the attractor. The remaining degrees of freedom in $\delta f$ are then all linearly independent from those of the attractor. 

For the typical case in which we require particular components of $I^{\lambda\mu_1\cdots\mu_n}$ to be carried entirely by the attractor piece $f_\mathcal{A}$, the matching conditions are already linear, and take the form
\begin{equation}
    M_J[f]=u_\lambda R^{(J)}_{\mu_1\cdots\mu_{n_J}} I^{\lambda\mu_1\cdots\mu_{n_J}}[f]\,. \label{eq:matching-form}
\end{equation}
Note that for the hydrodynamic currents $T^{\mu\nu}$ and $N^\mu$, this reproduces the usual Landau conditions with $R^{(J)}\in \{1,u^\nu\}$.

\paragraph{Viscous Attractodynamics}

The analog of dissipative hydrodynamics is obtained by retaining a finite set
of residual moments in $\delta f$, in addition to the dynamics of $\phi^I$.
One then needs to decide how many degrees of freedom will be kept in $\delta f$, by truncating some expansion in moments to a given desired accuracy.  Consequently, i) the 
equations of motion for $\phi^I$ are modified, and ii) new equations of motion need to be introduced 
for the residual degrees of freedom.

In this work we use the term ``viscous'' in this transient sense.  The residual variables are not set equal to their instantaneous gradient-expansion values, as in Navier--Stokes theory. Rather, as in Israel-Stewart hydrodynamics, they are independent degrees of freedom whose equations of motion are obtained by projecting the kinetic equation onto the retained residual directions. 

 Operationally, the simplest prescription is to take additional moments of the kinetic equation to provide the additional dynamical information that is needed. This is directly analogous to deriving Israel-Stewart hydrodynamics using the Grad 14-moment approximation~\cite{Grad:1949zza,Israel:1979wp}, but with the attractor $f_\mathcal{A}$ in place of an equilibrium distribution.

A useful attractodynamic truncation requires more than the existence of an
attracting manifold.  The off-attractor modes that are not retained must either
decay on timescales short compared with the evolution of the chart variables,
or be sourced only weakly over the regime of interest.  Equivalently, the
projection onto the retained chart and residual variables must be stable under
the subsequent evolution.  This separation need not be parametrically exact;
in the examples below it is a dynamical question that can be tested against
the full kinetic equation.

\subsection{This work: Proof of Principle Beyond Thermal Attractors} \label{sec:beyond-thermal}

While conceptually appealing, we emphasize that the preceding construction is not automatic: for a generic nonthermal
attractor it is not obvious that useful matching conditions and a controlled residual sector can be found. In the remaining sections of this work, we show that it is possible to carry out this construction in a specific kinetic theory featuring a non-thermal attractor. However, some of the properties that our derivation chiefly relies on can be formulated without reference to a specific collision kernel.

Given a kinetic theory with an attractor $f = f_{\mathcal{A}}(p;\phi^I)$, one can always consider expanding $f_{\mathcal{A}}$ around a suitable point\footnote{This depends on the kinematic range where the physics of interest lie; e.g., this might be the maximum of the distribution or some specific momentum scale whose dynamics one seeks to examine more closely.} as a function of $p$, such that
\begin{equation}
    \log f_{\mathcal{A}} = \alpha^{\mathcal{A}}(x) + \beta^{\mathcal{A}}_\mu(x) p^\mu + \frac12 p^\mu \mathcal{U}_{\mu \nu}^{\mathcal{A}}(x) p^\nu + \ldots \, , \label{eq:attractor-expansion}
\end{equation}
where the spacetime-dependent coefficients
$\alpha^\mathcal{A}$, $\beta^\mathcal{A}_\mu$, and
$\mathcal{U}^{\mathcal{A}}_{\mu\nu}$ are the corresponding coefficients of
$\log f_{\mathcal A}$ in the chosen momentum-space chart. They are therefore
functions of the attractor coordinates $\phi^I$, subject to the symmetries and
structural assumptions defining the attractor. In particular, not every
component need represent an independent attractor degree of freedom: some may
vanish, some may be fixed, and some may be related by the chosen symmetry
sector.

Therefore, given the dependence of the functions $\alpha^{\mathcal{A}},\beta_\mu^{\mathcal A}, \mathcal{U}_{\mu \nu}^{\mathcal A}$ on the attractor parameters $\phi^I$, the tangent directions to the attractor manifold are
\begin{equation}
    \frac{\partial f_{\mathcal A }}{\partial \phi^I} = \left( \frac{\partial \alpha^\mathcal{A} }{ \partial \phi^I} + p^\mu  \frac{\partial \beta_\mu^\mathcal{A} }{ \partial \phi^I} + \frac{1}{2} p^\mu p^\nu \frac{\partial \mathcal{U}_{\mu \nu}^\mathcal{A} }{\partial \phi^I} + \ldots \right) f_{\mathcal{A}} \, .\label{eq:tangent}
\end{equation}
This determines which perturbations lie parallel to the attractor manifold and which do not. This is crucial information once we promote the attractor parameters $\phi^I$ to general functions of spacetime, as their dynamics can source perturbations that lie both inside and outside of the attractor manifold, allowing for the possibility of coupling the attractor to viscous corrections.

We make a couple of comments regarding the properties that follow from the preceding discussion:
\begin{itemize}
    \item Eq.~\eqref{eq:tangent} is suggestive of the parametrization to choose for $\delta f$. In particular, it becomes convenient to choose the deviations from the attractor to contain 
    \begin{equation}
        \delta f \supset \left(w^{(0)} + w^{(1)}_\mu p^\mu + w^{(2)}_{\mu\nu} p^\mu p^\nu + \ldots \right) f_{\mathcal A} \, ,
    \end{equation}
    so that these (or some subset) may be fixed via matching conditions and not contain dynamics parallel to the attractor. 
    \item If the attractor is thermal ($\mathcal{U}^{\mathcal A}_{\mu \nu}=0$, $\beta^\mu\neq0$), the form of $\delta f$ in the preceding point is sufficient to describe some dissipative effects, as these may be accounted for by $w^{(2)}_{\mu\nu}$. However, if $\mathcal{U}^{\mathcal A}_{\mu \nu}\neq 0$, 
    then it may be natural to fix $w^{(2)}_{\mu\nu}$ via matching conditions and incorporate dissipative effects via additional terms, for example,
    \begin{equation}
        \delta f = \left(w^{(0)} + w^{(1)}_\mu p^\mu + w^{(2)}_{\mu\nu} p^\mu p^\nu + v^{(2)}_{\mu\nu} \frac{p^\mu p^\nu}{p\cdot u} + \ldots \right) f_{\mathcal A} \,.
    \end{equation}

\end{itemize}

We will see how these become manifest in a concrete example in what follows.

\section[BSY Attractodynamics in 0+1D]{BSY Attractodynamics in \(0+1\)D}
\label{sec:BSY}

We now demonstrate how the preceding construction can be carried out in practice. We do so by considering the BSY model~\cite{Brewer:2022vkq}, in which the attractor --- i.e., both the functional shape of the distribution function $f_{\mathcal{A}}$ and the attractor evolution equations --- is known analytically. This model is characterized by the kinetic equation
\begin{equation}
    \partial_\tau f -\frac{p_z}{\tau}\partial_{p_z} f = \lambda_0 \ell_{\rm Cb}[f] I_a[f]\nabla_p^2 f \, , \label{eq:kin-smallscatt}
\end{equation}
motivated from the small-angle scattering limit of the Boltzmann equation for gluons in weakly coupled QCD~\cite{Mueller:1999pi,Blaizot:2013lga}, under the assumption of boost-invariant profiles. Here $\ell_{\rm Cb}[f] = \ln (p_{\rm UV}/p_{\rm IR} )$ is the (perturbatively divergent) Coulomb logarithm coming from the integral over the small scattering angle, where $p_{\rm UV}$ and $ p_{\rm IR} $ are UV and IR cutoffs, respectively. We will use $p_{\rm UV} = \sqrt{\langle E p^2 \rangle / \langle E\rangle}$ and $p_{\rm IR} = m_D$, where $m_D^2 = 4 g_s^2 N_c \int \frac{d^3p}{(2\pi)^3} \frac{f}{p}$. The constant $\lambda_0$ is given by $\lambda_0 = 4\pi \alpha_s^2 N_c^2$, and $I_a[f] = \int \frac{d^3p}{(2\pi)^3} f (1+f)$ is the effective density of scatterers. For brevity we will henceforth use the notation $q[f]=\tau \lambda_0 \ell_{\rm Cb}[f] I_a[f]$. 

\subsection{Review of the attractor in the BSY model}

This theory is known to possess an exact attractor solution~\cite{Brewer:2022vkq} (its excited states in the Adiabatic Hydrodynamization framework are also known exactly~\cite{Brewer:2022vkq}, and the quasinormal modes of the nonthermal fixed point are also known~\cite{DeLescluze:2025gaa}). The functional form of the attractor is given by
\begin{equation}
    \ln(f_\mathcal{A})= \ln(A(\tau))-\frac12\left(\frac{p_\perp^2}{B^2(\tau)}+\frac{p_z^2}{C^2(\tau)}\right) \, ,
\end{equation}
and the attractor evolution equations are given by
\begin{subequations}
\label{eq:bsy}
\begin{align}
    \frac{\partial_y A}{A} &=  -2\frac{\partial_y B}{B} - \frac{\partial_y C}{C} -1 \, ,\\
    \frac{\partial_y B}{B} &= \frac{q}{B^2} \, , \\
    \frac{\partial_y C}{C} &= \frac{q}{C^2}-1 \, ,
\end{align}
\end{subequations}
where we are using $y = \ln (\tau/\tau_0)$. Here $q$, when evaluated on the attractor solution $f_{\mathcal A}$, is a function of $A,B$ and $C$. Faithful to the definition of an attractor, this is an exact solution of the kinetic equation~\eqref{eq:kin-smallscatt}, and is also stable under perturbations~\cite{Brewer:2022vkq,Mikheev:2022fdl,DeLescluze:2025gaa}. We shall refer to these attractor evolution equations as the ``BSY equations'' after the authors of~\cite{Brewer:2022vkq}.

In the notation of Eq.~\eqref{eq:attractor-expansion}, this attractor corresponds to
\begin{align} \label{eq:bsy-attractor-expansion}
    \alpha^\mathcal{A}= \ln(A(\tau)) \, , & & \beta_\mu^\mathcal{A}=0 \, , & &  \mathcal{U}_{\mu\nu}^{\mathcal{A}}=-\frac{1}{C^2}\ell_\mu\ell_\nu+\frac{1}{B^2}
    \Xi_{\mu\nu}
     \, ,
\end{align}
which could be interpreted as taking an infinite temperature limit (thus removing the term that would give rise to a Boltzmann distribution), preserving the Gaussian part, and enforcing axial symmetry about  $\ell$, coincident with the longitudinal direction associated with $p_z$ in Eq.~\eqref{eq:kin-smallscatt}. It would be interesting, but outside the scope of this work, to also allow $\ell^\mu$ to vary in a nontrivial way (e.g., as a way of accounting effects that break boost invariance). Therefore, in terms of the notation of the previous section, the chart variables $\phi^I$ of this attractor are $A,B$ and $C$. Equivalently, we will also use a dimensionless variable $\xi = (B/C)^2-1$ instead of $C$ interchangeably to parametrize the attractor.

Based on this exact attractor solution found by BSY~\cite{Brewer:2022vkq}, we will define and study both ideal attractodynamics (where only the attractor degrees of freedom appear) as well as the leading viscous corrections (where additional, dissipative equations appear). This can then be compared to evolution of the full kinetic theory in order to investigate the regime of validity of each attractodynamic theory.

\subsection{Attractodynamic and viscous degrees of freedom}

In order to construct an attractodynamic theory, once the attractor chart variables have been identified, we need to specify which ``macroscopic'' variables our theory will describe.

Just as in hydrodynamics, it is natural to consider the stress-energy tensor to furnish the low-energy description of the out-of-equilibrium dynamics, providing the energy and pressure(s) as macroscopic variables. However, as we discussed earlier in Section~\ref{sec:beyond-thermal}, due to the nature of the underlying attractor it will be natural to consider additional macroscopic quantities.

We start by discussing the number density and stress-energy tensor. We will assume that the system is even about $\ell^\mu$ and has rotational symmetry in the transverse plane.
As in aHydro~\cite{Romatschke:2003ms,Florkowski:2010cf,Martinez:2010sc,Martinez:2012tu,Bazow:2013ifa}, it is natural to write
\begin{align}
    T^{\mu\nu} &= T^{\mu\nu}_\mathcal{A} + \delta T^{\mu\nu}\\
    N^\mu & = N^\mu_\mathcal{A} + \delta N^\mu
\end{align}
where
\begin{align}
    T^{\mu\nu}_\mathcal{A} &= \mathcal{E}_\mathcal{A} u^\mu u^\nu - \mathcal{P}_{\perp,\mathcal{A}} \Xi^{\mu\nu} + \mathcal{P}_{L,\mathcal{A}} \ell^\mu \ell^\nu \\
    \delta T^{\mu\nu} &= \braket{E^2}_{\delta f} u^\mu u^\nu - \Pi \Delta^{\mu\nu} + \pi^{\mu\nu}\label{eq:dT}
\end{align}
and
\begin{align}
    N^\mu_\mathcal{A} &= n_\mathcal{A} u^\mu \\
    \delta N^\mu &= \braket{E}_{\delta f} u^\mu \,.
\end{align}
In analog to the expansion of $N^\mu, T^{\mu\nu}$ near equilibrium; here we instead expand near the attractor $f_\mathcal{A}$, meaning that here $\Pi$ and $\pi^{\mu\nu}$ differ from their usual hydrodynamic definition. Here the $\Pi$ and $\pi^{\mu\nu}$ terms in $\delta T^{\mu\nu}$ can be further decomposed in terms of $\Xi^{\mu\nu}$ and $\ell^\mu$, but here we have presented them in their more familiar form. We furthermore will follow the non-hydrodynamic current $I^{\lambda\mu\nu}$. Let us rewrite the tensor decomposition in Eq.~\eqref{eq:ilambdamunu-decomposition} in the form
\begin{align}
\begin{split}
    I^{\lambda\mu\nu} &= I_0 u^\lambda u^\mu u^\nu + I_\parallel (u^\lambda \ell^\mu \ell^\nu + u^\mu \ell^\lambda \ell^\nu + u^\nu \ell^\lambda \ell^\mu) \\
    &\qquad - I_{\perp} (u^\lambda \Xi^{\mu\nu} + u^\nu \Xi^{\lambda\mu} + u^\mu \Xi^{\lambda \nu} ) 
\end{split}
\end{align}

Because the particles in this theory are massless, the three macroscopic quantities we have defined in this current carry a total of two non-hydrodynamic degrees of freedom associated with the extra attractor degrees of freedom $B,C$. Then the division into attractor and fluctuation parts is
\begin{align}
&\begin{split}
    I_\mathcal{A}^{\lambda\mu\nu} &= I_{0,\mathcal{A}} u^\lambda u^\mu u^\nu  - I_{\perp,\mathcal{A}} (u^\lambda \Xi^{\mu\nu} + u^\nu \Xi^{\lambda\mu} + u^\mu \Xi^{\lambda \nu} ) \\
    &\qquad + I_{||,\mathcal{A}} (u^\lambda \ell^\mu \ell^\nu + u^\mu \ell^\lambda \ell^\nu + u^\nu \ell^\lambda \ell^\mu) \,. \\
\end{split}\\
&\begin{split}
    \delta I^{\lambda\mu\nu} &= \delta I_0 u^\lambda u^\mu u^\nu - \delta I_{\perp} (u^\lambda \Xi^{\mu\nu} + u^\nu \Xi^{\lambda\mu} + u^\mu \Xi^{\lambda \nu} ) \\
    &\qquad + \delta I_\parallel (u^\lambda \ell^\mu \ell^\nu + u^\mu \ell^\lambda \ell^\nu + u^\nu \ell^\lambda \ell^\mu) \,. \label{eq:deltai-decomposition}
\end{split}
\end{align}

The macroscopic attractor quantities with a subscript $\mathcal{A}$ in $N^\mu$, $T^{\mu\nu}$, and $I^{\lambda\mu\nu}$ can all be expressed in terms of the microscopic attractor variables $A,B,C$.

\paragraph{The generalized bulk and shear viscous pressures}

Note that the generalized bulk and shear viscous pressures $\Pi$ and $\pi^{\mu\nu}$ differ from thehydrodynamic quantities typically referred to by these symbols.  
Because we will proceed with an analog of the 14-moment approximation, the variables \(\Pi\) and \(\pi^{\mu\nu}\) should be understood as
retained residual moments, not as Navier--Stokes constitutive corrections.  In
the \(0+1\)D truncation below they are evolved dynamically, in direct analogy
with the transient shear and bulk sectors of Israel--Stewart theory~\cite{Israel:1979wp}.
We note that $\Pi$ and $\pi^{\mu\nu}$ are defined by
\begin{align}
    \Pi &= -\frac{1}{3}\left\langle \Delta_{\mu\nu}p^\mu p^\nu\right\rangle_{\delta f} \, , \\
    \pi^{\langle\mu\nu\rangle} &= \left\langle p^{\langle\mu} p^{\nu\rangle}\right\rangle_{\delta f} \, .
\end{align}
Furthermore, note that we have also written $\braket{E^2}_{\delta f} u^\mu u^\nu$ in Eq.~\eqref{eq:dT}, a term not present in either the hydro or aHydro expressions for $\delta T^{\mu\nu}$ due to the choice of the Landau matching condition $\braket{E^2}_{\delta f}=0$. 

Here, we will not impose this condition.  In hydrodynamics, Landau matching removes the energy-density component of the fluctuation and thereby fixes the inverse-temperature direction in the local equilibrium chart.  In a thermalizing attractodynamic chart, the analogous direction is
the linear \(\beta u\cdot p\) term.  The BSY attractor used in the present section is the special infinite-temperature scaling limit in which \(\beta_\mu^{\mathcal A}=0\), as in Eq.~\eqref{eq:bsy-attractor-expansion}. Since this reduced chart does not contain an energy/temperature direction, imposing \(\braket{E^2}_{\delta f}=0\) would amount to adding a field definition for a variable that is absent from the BSY truncation.  We therefore allow \(\braket{E^2}_{\delta f}\neq0\).  This is also natural in the BSY model because, due to one of its simplifying approximations, the diffusion kernel does not conserve energy exactly; the size of this effect is quantified in
Appendix~\ref{app:energy-non-conservation}.\footnote{We emphasize that this is a limitation of the BSY model, not of the attractodynamic construction itself. We also note that the BSY model \emph{does} conserve number.}  The timelike viscous contribution to \(T^{\mu\nu}\) is therefore not removed by matching and must be retained.
For massless particles, tracelessness of $\delta T^{\mu\nu}$ implies
\begin{equation}
    \braket{E^2}_{\delta f} = 3\Pi\,.
\end{equation}
In the simplified $0+1$D scenario we will consider here, $\pi^{\mu\nu}$ can be written in terms of a single independent scalar degree of freedom,
which we denote by $\pi$, such that
\begin{equation}
    \pi^{\mu\nu}
    =
    \mathrm{diag}\left(0,-\frac{\pi}{2},-\frac{\pi}{2},\pi\right)=\frac{\pi}{2}\left(3\ell^\mu\ell^\nu+\Delta^{\mu\nu}\right).
\end{equation}
in the local rest frame. Thus, we will have two dissipative degrees of freedom, $\pi$ and $\Pi$.

\paragraph{Matching conditions for the attractor variables}

We will still require matching conditions analogous to the Landau conditions; the equivalent statement in our case should ensure that fluctuations are perpendicular to the attractor manifold parametrized by $A$, $B$, and $C$. This gives us
\begin{equation}\label{eq:newmatching}
    \braket{E}_{\delta f}=\braket{Ep_z^2}_{\delta f} = \braket{Ep_\perp^2}_{\delta f} = 0\,,
\end{equation}
which will make $A$, $B$, and $C$ well-defined and restrict $\delta f$ as being orthogonal to the attractor manifold. These matching conditions also fully fix $I^{\lambda\mu\nu}$. In the language of the decomposition in Eq.~\eqref{eq:deltai-decomposition}, the latter two matching conditions are
\begin{equation}
    \delta I_\parallel = \delta I_{\perp} = 0\,,
\end{equation}
and in the language of Eq.~\eqref{eq:matching-form}, they are
\begin{align}
    u_\lambda \ell_\mu \ell_\nu I^{\lambda \mu \nu}[f] &= u_\lambda \ell_\mu \ell_\nu I^{\lambda \mu \nu}[f_\mathcal{A}]\,,\\
    u_\lambda \Xi_{\mu\nu} I^{\lambda \mu \nu}[f] &= u_\lambda \Xi_{\mu\nu} I^{\lambda \mu \nu}[f_\mathcal{A}]\,.
\end{align}

We will proceed by analogy with the 14-moment approximation, in a way that naturally encodes our additional degrees of freedom. First, we review the Grad 14-moment approximation with Boltzmann statistics~\cite{Grad:1949zza}. Traditionally, one writes
\begin{equation}
    \delta f = f_0 \left(\alpha + \beta_\mu p^\mu + w_{\mu\nu} p^\mu p^\nu\right)\,.
\end{equation}
The equilibrium distribution in this case is of the form $\exp(a+b p)$; one therefore notes that $\alpha$ and $\beta_0$ describe fluctuations tangent to the equilibrium manifold,
\begin{equation}
    \frac{\partial f_0}{\partial a} = f_0 \quad\quad \frac{\partial f_0}{\partial b} = p f_0\,.
\end{equation}
As we have already mentioned, this necessitates the use of matching conditions to clarify which part of $f$ belongs in $f_0$, and which belongs in $\delta f$ (thereby making $a$ and $b$ well-defined). The usual Landau matching conditions require that $\delta f$ carries no number or energy density, that is, for a gas of relativistic particles, $\braket{E}_{\delta f}=\braket{E^2}_{\delta f}=0$. We note the direct connection to the directions tangent to the equilibrium manifold above. In the sense of Eq.~\eqref{eq:associated-moments}, $\alpha$ corresponds to the chemical potential $\mu$ and $\beta_\mu u^\mu$ corresponds to the temperature $T$ via the Landau matching conditions, while $w_{\mu\nu}$ describes non-trivial dissipative currents in the $0+1$D theory.

In our theory, the equivalent of $a,b$ are the coordinate rescalings $A,B,C$ which parameterize our attractor surface. As we discussed in Section~\ref{sec:beyond-thermal}, we can write the directions tangent to this manifold as
\begin{equation} \label{eq:bsytangent}
    \frac{\partial\ln(f_\mathcal{A})}{\partial\ln(A)}=1,\qquad\frac{\partial\ln(f_\mathcal{A})}{\partial\frac{1}{B^2}}=-\frac{1}{2}p_\perp^2,\qquad\frac{\partial\ln(f_\mathcal{A})}{\partial\frac{1}{C^2}}=-\frac{1}{2}p_z^2.
\end{equation}
We want to consider an ansatz for $\delta f$ which naturally captures both our attractor and dissipative degrees of freedom. A natural expression to write is
\begin{equation}
    \delta f = f_\mathcal{A} \left(\alpha + \beta_\mu p^\mu + w_{\mu\nu} p^\mu p^\nu + v_{\mu\nu} \frac{p^\mu p^\nu}{p\cdot u} \right)\,.
\end{equation}
where now each term in $\delta f$ corresponds to a moment we calculate in $T^{\mu\nu}$, $N^{\mu}$ , or our matching conditions. In $0+1$D --- assuming reflection symmetry about the longitudinal axis --- this reduces to
\begin{equation} \label{eq:simplified-ansatz}
    \delta f = f_\mathcal{A} \left(\alpha + \beta_0 \zeta + w_{00} \zeta^2 + w_{33} \zeta^2 \pzp^2 + v_{33} \zeta \pzp^2 \right)
\end{equation}
where $\pzp = p_z/p$, $\zeta=p/B$, and we have rescaled the definitions of the latter coefficients for convenience.
In the same loose sense as $(\alpha,\beta_0)$ corresponded to $(\mu,T)$ in the preceding example, the constant $\alpha$ corresponds to $A$ via the matching condition for number density, $w_{00}$ and $w_{33}$ correspond to $B$ and $C$ via the matching conditions Eq.~\eqref{eq:newmatching}, $\beta_0$ corresponds to $\Pi$, and $v_{33}$ corresponds to $\pi$. We therefore use the matching conditions plus the definitions of $\pi$ and $\Pi$ to eliminate the coefficients $\alpha,\beta_0, w_{00},w_{33},v_{33}$ in favor of the attractodynamic quantities $A,B,C,\pi,\Pi$.

\subsection{Dynamics}\label{sec:dynamics}

Writing the distribution function as $f=f_\mathcal{A}+\delta f$, the Boltzmann equation is
\begin{equation}
    \partial_y f_\mathcal{A} + \partial_y \delta f= p_z\partial_{p_z} f_\mathcal{A} +p_z\partial_{p_z} \delta f +q[f]\nabla_p^2 f_\mathcal{A} + q[f]\nabla_p^2 \delta f 
\end{equation}
We can group the terms in the Boltzmann equation into two categories, classified by whether they are tangent or orthogonal to the attractor manifold,
\begin{align}
     \partial_y f_\mathcal{A}&= p_z\partial_{p_z} f_\mathcal{A} + q[f]\nabla_p^2 f_\mathcal{A} \label{eq:attractoreom}\\
     \partial_y \delta f&= p_z\partial_{p_z}\delta f + q[f]\nabla_p^2 \delta f \label{eq:dfeom}\,,
\end{align}
where here we have identified the orthogonal pieces as those which vanish under the operations
\begin{equation}\label{eq:bsymoments}
    \int \frac{d^3p}{(2\pi)^3} p_z^2 \,\bm{\cdot}\;, \quad\quad \int \frac{d^3p}{(2\pi)^3} p_\perp^2 \,\bm{\cdot}\;, \quad\quad \int \frac{d^3p}{(2\pi)^3} \,\bm{\cdot}
\end{equation}
based on Eq.~\eqref{eq:bsytangent}. Note that \textit{we have not linearized the Boltzmann equation}, although one may equally derive this theory of attractodynamics from the linearized kinetic theory, as we show in Appendix~\ref{app:linearized}. The moments~\eqref{eq:bsymoments} of Eq.~\eqref{eq:attractoreom} will give the viscous version of the BSY equations, which will differ only through the contribution of $\delta f$ to $q$. Moments of Eq.~\eqref{eq:dfeom} will give equations of motion for the viscous pressures according to
\begin{align}
    \partial_y \pi &= \int \frac{d^3p}{(2\pi)^3} p \left(\pzp^2 - \frac{1}{3} \right)  \partial_y \delta f\,,\label{eq:dypi}\\
    \partial_y \Pi &= \frac{1}{3} \int \frac{d^3p}{(2\pi)^3} p\,  \partial_y \delta f\,.\label{eq:dyPi}
\end{align}

\paragraph{Explicit form of the viscous evolution equations}

In practice, it will be more convenient to write the attractor distribution function in polar coordinates for the purpose of computing these moments; that is,
\begin{equation}
    \ln(f_\mathcal{A})= \ln(A(\tau))-\frac12\frac{p^2}{B^2(\tau)}\left(1+\xi \pzp^2\right),
\end{equation}
where we have defined $\xi=(B/C)^2-1$. We note that $\xi, B$ are directly analogous to $\xi,\Lambda$ in the traditional Romatschke-Strickland form~\cite{Romatschke:2003ms} of the distribution function in anisotropic hydrodynamics. (We are considering a massless theory with $p^2=p_\perp^2+p_z^2$.) We then replace the equation of motion for $C(\tau)$ with
\begin{equation}
    \partial_y \xi = 2(\xi+1)\left(1-\xi\frac{q}{B^2}\right)\,.
\end{equation}

We organize the integrals by defining, along with $w_\mathcal{A} = f_\mathcal{A}/A$, 
\begin{align}
    \bm{c} &= (\alpha, \beta_0, w_{00}, w_{33}, v_{33})\\
    \bm{g} &= (1, \zeta, \zeta^2, \pzp^2\zeta^2, \pzp^2\zeta) w_\mathcal{A}\\
    \mathcal{O}_1 &= \pzp^2 \zeta \partial_\zeta + \pzp (1-\pzp^2)\partial_\pzp\\
    \mathcal{O}_2 &= \frac{1}{\zeta^2} \left[ \partial_\zeta (\zeta^2 \partial_\zeta\, \bm{\cdot} ) + \partial_\pzp ((1-\pzp^2) \partial_\pzp\, \bm{\cdot}) \right]\\
    \mathcal{I}^{(\pi)}_{ij} &= \frac{1}{(2\pi)^2} \int d\zeta\, d\pzp \zeta^3 \left(\pzp^2 -\frac{1}{3}\right) \mathcal{O}_i g_j\\
    \mathcal{I}^{(\Pi)}_{ij} &= \frac{1}{(2\pi)^2} \int d\zeta\, d\pzp \zeta^3 \left(\frac{1}{3}\right) \mathcal{O}_i g_j\,,
\end{align}
where the integrals $\mathcal{I}^{(\pi)}_{ij},\mathcal{I}^{(\Pi)}_{ij}$ are simple to compute and are functions of $\xi$ only. then Eqs.~\eqref{eq:dypi} and~\eqref{eq:dyPi} can be written
\begin{align}
    \partial_y \pi &= AB^4 \left[ \mathcal{I}^{(\pi)}_{1i} + \frac{q}{B^2} \mathcal{I}^{(\pi)}_{2i} \right] c_i\,,  \label{eq:dypi-explicit}\\
    \partial_y \Pi &= AB^4 \left[ \mathcal{I}^{(\Pi)}_{1i} + \frac{q}{B^2} \mathcal{I}^{(\Pi)}_{2i} \right] c_i  \label{eq:dyPi-explicit}\,.
\end{align}
The functional $q$ can also be written in such a way that all integrals are computed in advance. We define
\begin{align}
    \tilde{\bm{c}} &= (1, \bm{c})\,, \\
    \tilde{\bm{g}} &= (w_\mathcal{A}, \bm{g})\,,
\end{align}
and will require the integrals
\begin{align}
    \tilde{\mathcal{J}}^{(n)}_i &= \frac{1}{(2\pi)^2} \int d\zeta\, d\pzp \zeta^{(2+n)} \tilde{g}_i\,, \\
    \tilde{\mathcal{J}}_{ij} &= \frac{1}{(2\pi)^2} \int d\zeta\, d\pzp \zeta^2 \tilde{g}_i \tilde{g}_j\,,
\end{align}
which will again be functions of $\xi$ only. Now the relevant functionals can be written
\begin{align}
    I_a &= AB^3 \tilde{c}_i \tilde{\mathcal{J}}^{(0)}_i + A^2 B^3 \tilde{c}_i \tilde{c}_j \tilde{\mathcal{J}}_{ij} \\
    m_D^2 &= 4 g_s^2 N_c AB^2 \tilde{c}_i \tilde{\mathcal{J}}^{(-1)}_i \\
    p_{UV}^2 &= \frac{B^2 \tilde{c}_i \tilde{\mathcal{J}}^{(2)}_i}{\tilde{c}_j \tilde{\mathcal{J}}^{(0)}_j} 
\end{align}
where we remind the reader that $q = \lambda_0 \tau \ell_{\rm Cb} I_a$.

\subsection{Attractodynamics in action}\label{sec:numerics}

We will now compare ideal attractodynamics, viscous attractodynamics, and a full numerical solution to the Boltzmann equation in the BSY model. This is a verification that there exists a regime in which a given attractodynamic truncation of kinetic theory provides a consistent description, despite following only a small, finite number of degrees of freedom, compared to the effectively infinite degrees of freedom of a distribution function.

\begin{figure}
    \centering
    \includegraphics[width=0.48\linewidth]{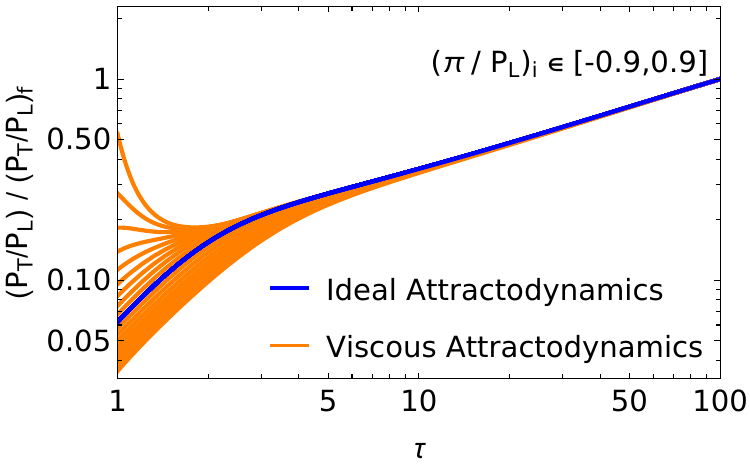}
    \includegraphics[width=0.48\linewidth]{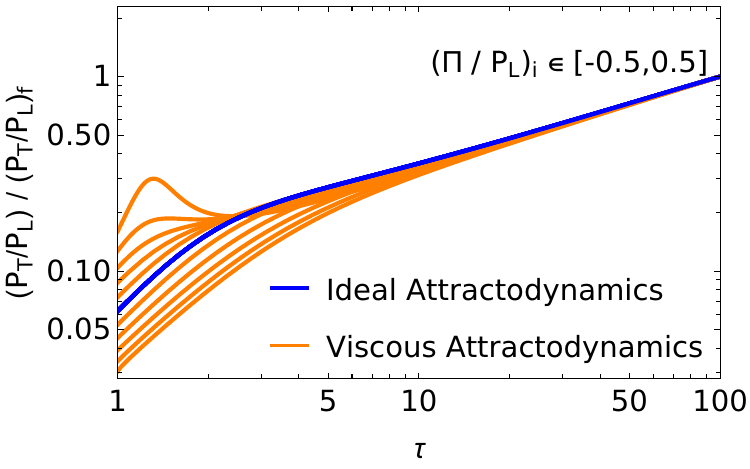}
    \caption{A comparison between ideal and viscous BSY attractodynamics, investigating the breakdown as initial conditions diverge from the attractor manifold. Different orange lines correspond to equally spaced choices of initial conditions for $\pi$ ($\Pi$) in the left (right) plot.}
    \label{fig:attracts}
\end{figure}

We choose a coupling of $g_s=0.1$ and an attractodynamic initial condition of $A_0=\sigma_0/g_s^2$, $B_0=1/\sqrt{2}$, and $\xi_0=4$ with $\sigma_0=4$ for the purpose of demonstration; other values are considered in Appendix~\ref{app:scan-initial-conditions}. This is sufficient to initialize ideal attractodynamics, which we evolve according to the BSY equations~\eqref{eq:bsy} with $q=q[f_\mathcal{A}]$. We can additionally specify an initial value of $\pi$ and $\Pi$, which will complete the initialization of viscous attractodynamics. In the viscous case, we evolve Eq.~\eqref{eq:bsy} with $q=q[f]$, plus equations of motion for $\pi$ and $\Pi$ as derived in Sec.~\ref{sec:dynamics}. In Fig.~\ref{fig:attracts}, we scan possible initial values of these viscous generalized pressures, and observe the expected attractor behavior.

\begin{figure}
    \centering
    \includegraphics[width=0.48\linewidth]{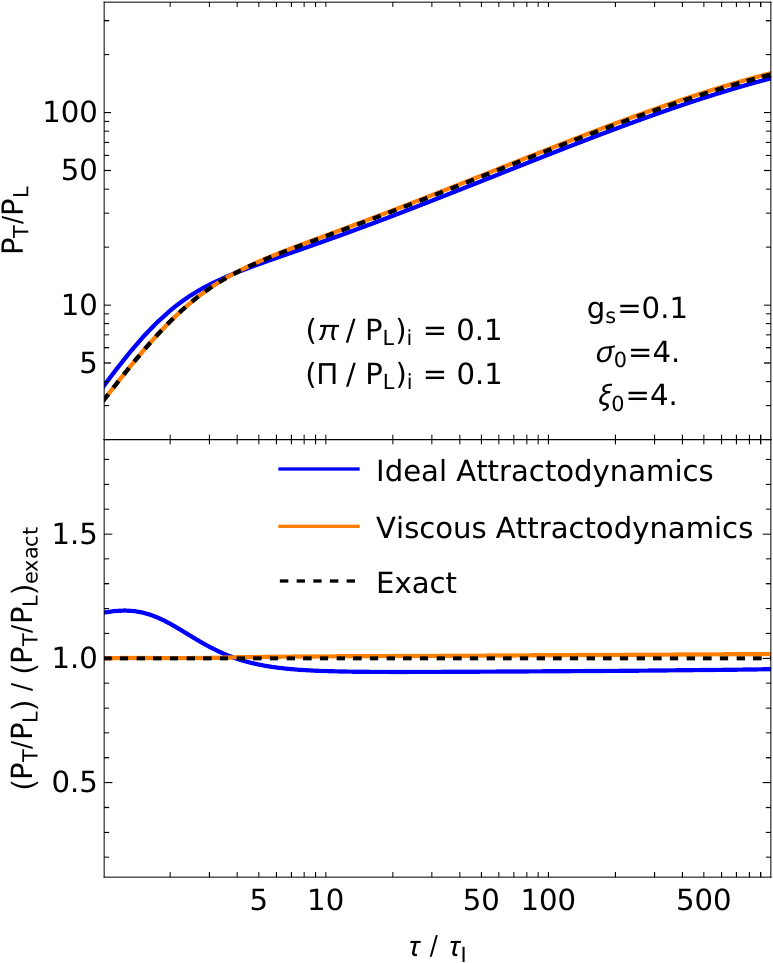}
    \includegraphics[width=0.48\linewidth]{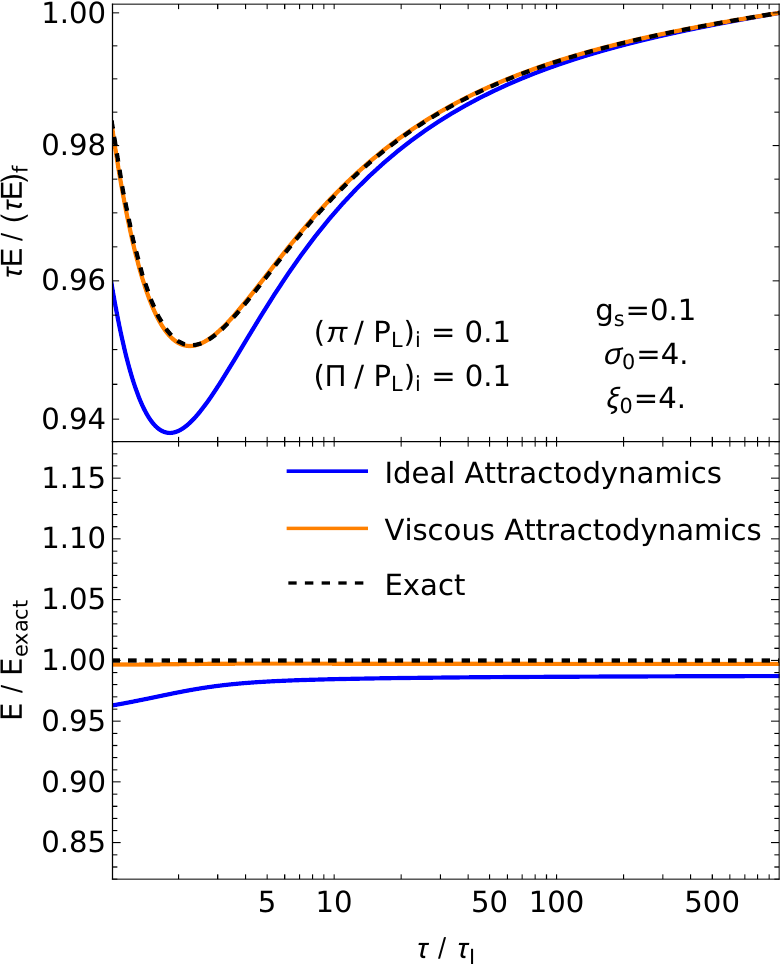}
    \caption{A comparison between attractodynamics and a complete numerical solution of the Boltzmann equation in pressure anisotropy (left) and energy density (right). For the viscous case, we choose $\pi_0/(\mathcal{P}_L)_i = \Pi_0/(\mathcal{P}_L)_i = 0.1$, and use our matching conditions to write the corresponding form of $f=f_\mathcal{A}+\delta f$ as the initial condition for the full numerical solution.}
    \label{fig:fullthy}
\end{figure}

For the full numerical solution, we initialize $f$ according to the ansatz~\eqref{eq:simplified-ansatz} with coefficients specified by our matching procedure given $A_0,B_0,\xi_0,\pi_0,\Pi_0$. Thus the exact and viscous solutions will agree exactly at the initial time $\tau_I=1$ by construction. We can then check that for an initial condition close to the attractor surface (we choose $\pi_0/(\mathcal{P}_L)_i = \Pi_0/(\mathcal{P}_L)_i = 0.1$ in Fig.~\ref{fig:fullthy}), attractodynamics provides a very good description, and viscous attractodynamics outperforms ideal attractodynamics. Notably, although the exact theory eventually falls onto the attractor surface ($\pi,\,\Pi \rightarrow 0$), the ideal attractodynamics calculation does not agree with the exact and viscous solutions at late times. This is because the initial viscous generalized pressures affect the value of $q$ in the BSY equations, meaning that the system is driven to a different part of the attractor surface.

\begin{figure}
    \centering
    \includegraphics[width=0.48\linewidth]{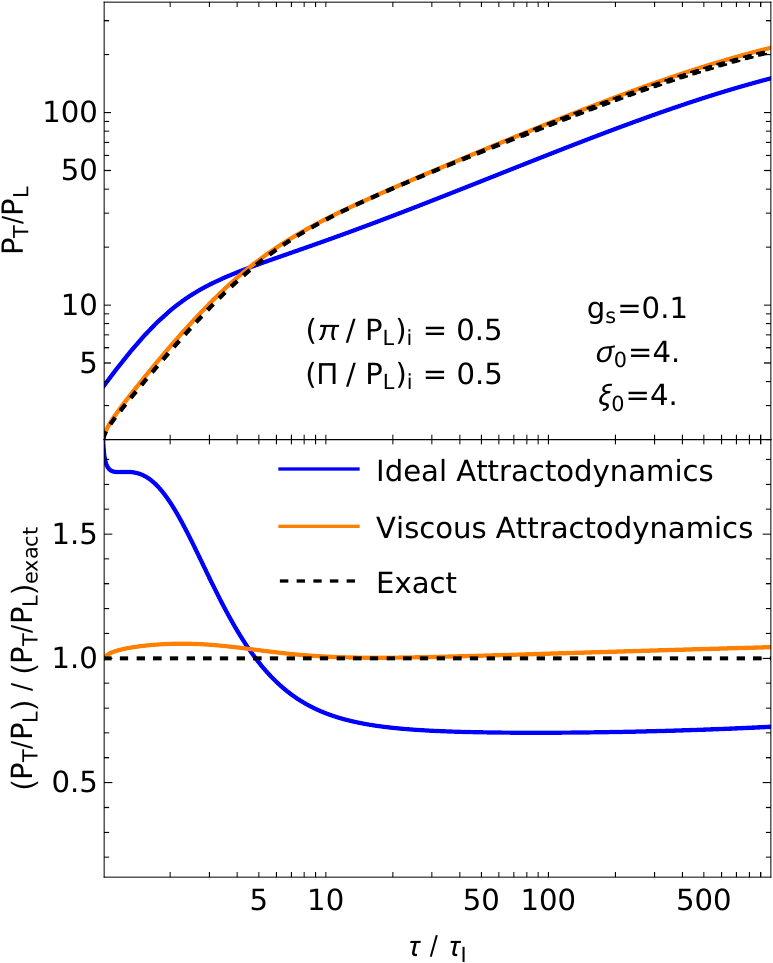}
    \includegraphics[width=0.48\linewidth]{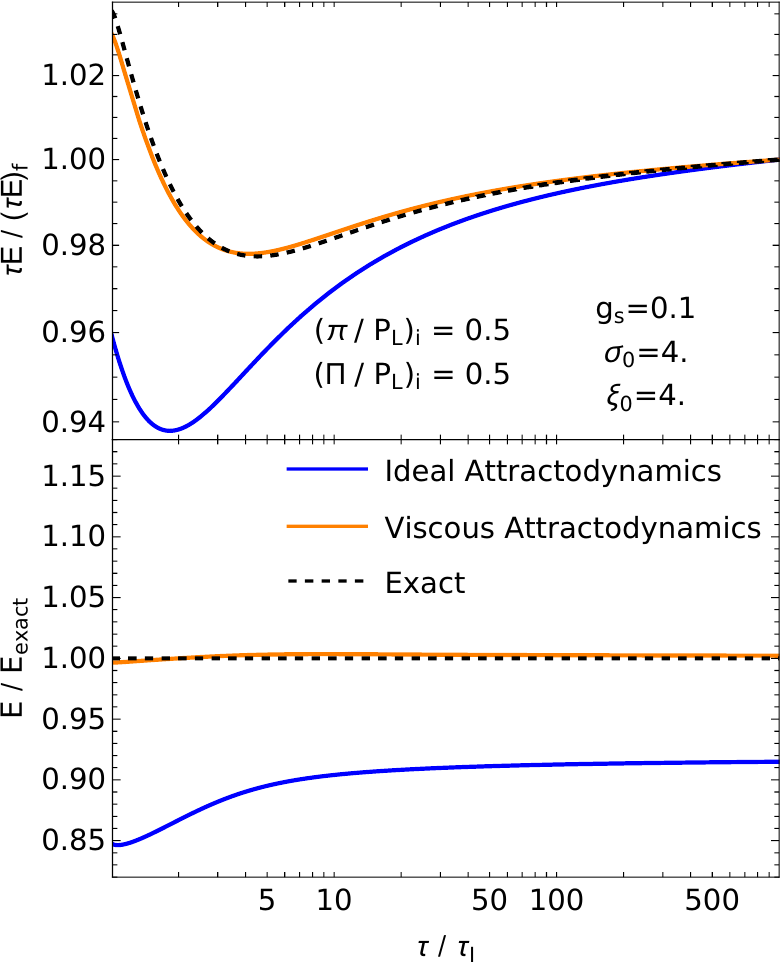}
    \caption{As in Fig.~\ref{fig:fullthy}, but with a larger deviation from the attractor surface $\pi_0/(\mathcal{P}_L)_i = \Pi_0/(\mathcal{P}_L)_i = 0.5$.}
    \label{fig:bigpi}
\end{figure}

Furthermore, we can check the performance of viscous attractodynamics as we increase the deviation from the attractor manifold. In Fig.~\ref{fig:bigpi}, we use  $\pi_0/(\mathcal{P}_L)_i = \Pi_0/(\mathcal{P}_L)_i = 0.5$, and note that the performance of both ideal and viscous attractodynamics is worse, as expected.

\section{Attractodynamics Beyond Kinetic Theory} \label{sec:attract-beyond}
We now preempt how one might construct an attractodynamic theory from a macroscopic point of view, having seen what structures arise naturally when deriving one from a kinetic theory with a known attractor. We saw that the (local) symmetries of the attractor allowed a systematic organization of the macroscopic degrees of freedom. In the kinetic theory realization the assumption that the distribution function approaches the attractor allows the organization of the attractor degrees of freedom into moments of the distribution function. From the macroscopic perspective there is no longer a distribution function to take moments of. Nevertheless, it seems natural to encode the attractor degrees of freedom in general into a hierarchy of moment-like (maximally symmetric in indices) tensors\footnote{If the attractor has long lived high momentum modes in the microscopic theory (such that the attractor is not well-captured by a small momentum expansion, but nonetheless has a finite number of degrees of freedom), one would have to leverage more details of the attractor to organize the equations.}
\begin{align}
    N^\mu \, , & &
    T^{\mu \nu} \, , &
    &I_{(2)}^{\mu \nu_1\nu_2} \, , &
    &\cdots \, , &
    &I_{(n)}^{\mu \nu_1\cdots\nu_n} \, .
\end{align}
Using the representations of the symmetries of the attractor (and the local tensor structures specifying such symmetries; in our case, in terms of $u$ and $\ell$), one can decompose each of these tensors into the dynamical (scalar) degrees of freedom. For example, for an axisymmetric attractor, as we considered previously, described locally by $u^\mu(x),
\ell^\mu(x)$ with $u\cdot\ell=0$ we can decompose 
\begin{align}
    &N^\mu=n u^\mu+\tilde{n}\ell^\mu\\
    &T^{\mu \nu}= \mathcal{E} u^\mu u^\nu - \mathcal{P}_\perp \Xi^{\mu\nu} + \mathcal{P}_L \ell^\mu \ell^\nu \\
    &I_{(2)}^{\mu \nu_1\nu_2}= I_0 u^{(\mu} u^{\nu_1} u^{\nu_2 )} -3 I_\perp  u^{(\mu}\Xi^{\nu_1\nu_2)} + 3 I_\parallel  u^{(\mu}\ell^{\nu_1}\ell^{\nu_2)}  \,.
\end{align}
We then expect evolution equations of the form
\begin{equation}
    \partial_\mu I_{(n)}^{\mu \nu_1\cdots\nu_n}=-\mathcal{C}_{(n)}^{ \nu_1\cdots\nu_n},
\end{equation}
such that, again, only scalar quantities living in $u\cdot I_{(n)}$ (densities) are evolved by the equations of motion, in analogy to energy and number density, while the rest are pressure-like quantities (fluxes) requiring constitutive relation specifying them in terms of the densities. The source tensors $\mathcal{C}_{(n)}^{ \nu_1\cdots\nu_n}$ can also be expanded using representation theory, yielding sources that are in general functions of the densities, that is another needed form of constitutive input either from measurement or a microscopic theory. 

We emphasize here that the requirement of the additional tensors $I_{(n)}$ with $n\geq 2$ is due to the additional microscopic degrees of freedom included in the attractor surface as compared to the thermal attractor. Unlike $N^{\mu}$ and $T^{\mu\nu}$, these higher moments are not restricted by symmetry to be sourceless. This is where the increased sensitivity of the far-from-equilibrium dynamics on the microscopic details is captured; here we interpret that additional sensitivity as encoding the dynamics of the attractor.
In hydrodynamics, these degrees of freedom would appear at higher orders via transport coefficients as corrections to $T^{\mu\nu}$. In aHydro \cite{Strickland:2017kux} this is interpreted as the attractor degrees of freedom re-summing an infinite number of transport coefficients in inverse Reynolds number.

In hydrodynamics, one often finds that working in terms of thermodynamic quantities (such as $\mu, T$), instead of purely in terms of the macroscopic quantities (such as $n, \mathcal{E}$). In the kinetic theory setting this is exactly what matching conditions make manifest. Similarly, one might more generically in attractodynamics find it useful to define a mapping between coordinates on a (now abstract) attractor manifold and the macroscopic densities and fluxes, where there is a one-to-one mapping between the manifold coordinates and the macroscopic densities, while the fluxes as a function of the coordinates can encode the generalized equations of state. Depending on the symmetries of the attractor, the kinetic theory expansion of the attractor distribution function in Eq.~\eqref{eq:attractor-expansion} can provide a natural example of such a choice of coordinates.

Following the usual macroscopic approach to the gradient expansion in hydrodynamics, one can expand each given tensor $I_{(n)}$ in gradients of the relevant quantities.
A more in depth exploration of the details and peculiarities of this structure in general macroscopic attractodynamics is warranted, but outside of the scope of this work.

\section{Conclusions and Outlook}
\label{sec:conclusions}

Hydrodynamics is powerful not merely because local thermal equilibrium is a late-time attractor, but because it provides a closed macroscopic theory of long-wavelength dynamics near that attractor.  Macroscopic hydrodynamics is defined by its local variables, its conservation laws, and its constitutive relations, while microscopic information enters through the equation of state and transport coefficients.  In this work we asked whether an analogous macroscopic closure can be built around a far-from-equilibrium nonthermal attractor.  The central claim is that this is possible: near-attractor dynamics can be organized in terms of a finite set of macroscopic variables, generalized constitutive data, source terms, and, when needed, transient viscous modes. Because the additional moment-like variables relevant to a nonthermal attractor are not protected by exact conservation laws, their source terms carry more theory-dependent microscopic information than in ordinary hydrodynamics, but in an organized macroscopic form.

We demonstrated this idea for the first time, using the example of a kinetic theory, where the decomposition 
\begin{equation}
    f=f_{\mathcal A}+\delta f
\end{equation}
can be made explicit, and the corresponding macroscopic variables, matching conditions, evolution equations, and viscous truncation can be derived in a method analogous to the 14-moment approximation.  The matching conditions fix which part of a nearby distribution is assigned to the attractor chart and which part is treated as residual.  The ideal attractodynamic theory evolves only the chart variables.  Viscous attractodynamics instead retains selected off-attractor moments as independent viscous variables, in direct analogy with the role of nonhydrodynamic modes in Israel--Stewart-type extensions of hydrodynamics.  This kinetic realization is not meant to define attractodynamics itself; rather, it provides a controlled setting in which the macroscopic variables, matching prescription, projected equations, and viscous sector can be constructed and tested explicitly.

The BSY model provides such a controlled proof of principle.  Its $\beta=0$ Gaussian scaling solution is an exact nonthermal attractor, so the chart variables, tangent directions, and matching conditions can be identified without ambiguity or needing approximation.  In this example the chart is parameterized by $A,B,C$, while the leading viscous sector is described by two generalized viscous pressures, $\Pi$ and $\pi$.  Projecting the kinetic equation then gives both the ideal BSY attractodynamic equations and their leading viscous extension.  Comparison with the full kinetic equation shows the expected pattern: ideal attractodynamics describes evolution initialized on or close to the attractor, the viscous extension improves the description for nearby off-attractor initial data, and sufficiently large residual perturbations reveal the breakdown of the finite truncation.

Several simplifications were helpful in the present construction.  The BSY attractor is known exactly, we work in $0+1$D, the anisotropy direction is fixed by boost-invariant symmetry, and the dynamics reduces to ordinary differential equations.  In addition, the diffusion kernel used in the BSY model is not an exactly energy-conserving kinetic theory, a limitation quantified in Appendix~\ref{app:energy-non-conservation}.  The result should therefore be viewed as a controlled $0+1$D proof of principle rather than a complete local $3+1$D attractodynamic theory.

Looking forward, a central task is to turn attractodynamics from a kinetic construction into a standalone macroscopic framework. The BSY Gaussian attractor should not be viewed as universal or as a required input in future applications; rather, different systems may have different attractor symmetries, leading to different low-order charts, moment-like tensors, source terms, and viscous completions. In such a formulation, the theory-dependent input would be encoded in a finite set of constitutive functions, source terms, and residual relaxation data.  That input could be obtained from an explicit microscopic description, such as a kinetic collision kernel, but it need not be.  As in hydrodynamics, the closure data could also be modeled, constrained by microscopic calculations, or extracted phenomenologically. 
In heavy-ion applications, a longer-term goal would be to construct local attractodynamic closures constrained by microscopic calculations of pre-equilibrium attractor dynamics and by the late-time hydrodynamic limit, and then use Bayesian calibration to determine which of their remaining theory-dependent source and relaxation data are constrained by observables.  Adiabatic Hydrodynamization~\cite{Rajagopal:2024lou,Rajagopal:2025nca} may help identify the long-lived attractor modes that such a closure should retain in the weakly coupled regime.

At the same time, a standalone attractodynamic theory requires a sharper mathematical foundation.  The local initial-value problem must be well posed, the admissible local symmetry data (such as $u^\mu$ and $\ell^\mu$) and charts must be characterized, and the residual spectrum must be organized in a way that makes the truncation systematically improvable.  These questions are not specific to heavy-ion collisions.  Analogous constructions may be useful in other systems with far-from-equilibrium attractor behavior, including cold atomic gases, cosmological defect networks, and more general non-equilibrium quantum field theories.  The broader goal is therefore not only to identify far-from-equilibrium attractor solutions, but to turn them into organizing principles for more general macroscopic far-from-equilibrium dynamics.  The BSY example studied here is a concrete first controlled step towards such a construction.

\acknowledgments
We gratefully acknowledge helpful conversations with Florian Lindenbauer and Krishna Rajagopal.
Research supported in part by the U.S.~Department of Energy, Office of Science, Office of Nuclear Physics under grant Contract Number DE-SC0011090, by grant NSF PHY-2309135 to the Kavli Institute for Theoretical Physics (KITP), and by grant 994312 from the Simons Foundation.

\appendix

\section{Linearized BSY Attractodynamics}\label{app:linearized}

The linearized Boltzmann equation is 
\begin{equation}
    \partial_y f_\mathcal{A} + \partial_y \delta f= p_z\partial_{p_z} f_\mathcal{A} +p_z\partial_{p_z} \delta f +(q[f_\mathcal{A}]+\delta q[f_\mathcal{A},\delta f])\nabla_p^2 f_\mathcal{A} + q[f_\mathcal{A}]\nabla_p^2 \delta f
\end{equation}
where we can write $\delta q$ as a sum of two pieces,
\begin{align}
    \delta q[f_\mathcal{A},\delta f] &= \tau \lambda_0 I_a [f_\mathcal{A}] \delta \ell_{\rm Cb} [f_\mathcal{A},\delta f] + \tau \lambda_0 \delta I_a [f_\mathcal{A},\delta f] \ell_{\rm Cb} [f_\mathcal{A}]\,.
\end{align}
For the first piece, we have
\begin{align}
    \delta \ell_{\rm Cb} &= \frac{1}{2p_{\rm UV}^2 [f_\mathcal{A}]} \delta p_{\rm UV}^2 - \frac{1}{2 p_{\rm IR}^2 [f_\mathcal{A}]} \delta p_{\rm IR}^2
\end{align}
where 
\begin{align}
    \delta p_{\rm UV}^2 &= \frac{1}{\int \frac{d^3p}{(2\pi)^3} f_\mathcal{A} } \left[ \int \frac{d^3p}{(2\pi)^3} p^2 \delta f - p_{\rm UV}^2 [f_\mathcal{A}] \int \frac{d^3p}{(2\pi)^3} \delta f  \right] \\
    \delta p_{\rm IR}^2 &= 4 g_s^2 N_c \int \frac{d^3p}{(2\pi)^3} \frac{\delta f}{p}\,.
\end{align}
For the second piece, we have
\begin{align}
    \delta I_a &= \int \frac{d^3 p}{(2\pi)^3} (\delta f + 2 f_\mathcal{A} \delta f)\,.
\end{align}
Defining
\begin{align}
    \mathcal{J}^{(n)}_i &= \frac{1}{(2\pi)^2} \int d\zeta\, d\pzp \zeta^{(2+n)} g_i \\
    \mathcal{J}_{\mathcal{A} i} &= \frac{1}{(2\pi)^2} \int d\zeta\, d\pzp \zeta^2 w_\mathcal{A} g_i \,,
\end{align}
we can use our ansatz for $\delta f$ to compute all integrals in advance as in Sec.~\ref{sec:dynamics},
\begin{align}
    \delta p_{\rm UV}^2 &= (2\pi)^{3/2} B^2 c_i \sqrt{1+\xi} \left[ \mathcal{J}_i^{(2)} - \left(\frac{3+2\xi}{1+\xi}\right) \mathcal{J}_i^{(0)}  \right] \\
    \delta p_{\rm IR}^2 &= 4 g_s^2 N_c AB^2 c_i \mathcal{J}_i^{(-1)}\\
    \delta I_a &= AB^3 c_i \mathcal{J}_i^{(0)} + 2 A^2 B^3 c_i \mathcal{J}_{\mathcal{A} i}\,.
\end{align}

In the linearized theory, the equations of motion equivalent to Eqs.~\eqref{eq:attractoreom} and~\eqref{eq:dfeom} are
\begin{align}
     \partial_y f_\mathcal{A}&= p_z\partial_{p_z} f_\mathcal{A} + (q[f_\mathcal{A}]+\delta q)\nabla_p^2 f_\mathcal{A} \\
     \partial_y \delta f&= p_z\partial_{p_z}\delta f + q[f_\mathcal{A}]\nabla_p^2 \delta f \,.
\end{align}
From this we can see that the viscous terms $\pi$ and $\Pi$ can be computed in exactly the same way as in the main text, merely replacing the $q$ in Eqs.~\eqref{eq:dypi-explicit} and~\eqref{eq:dyPi-explicit} with $q[f_\mathcal{A}]$. Furthermore, the $q$ in the BSY equations of motion~\eqref{eq:bsy} will be replaced by $q[f_\mathcal{A}]+\delta q$.

\begin{figure}
    \centering
    \includegraphics[width=0.48\linewidth]{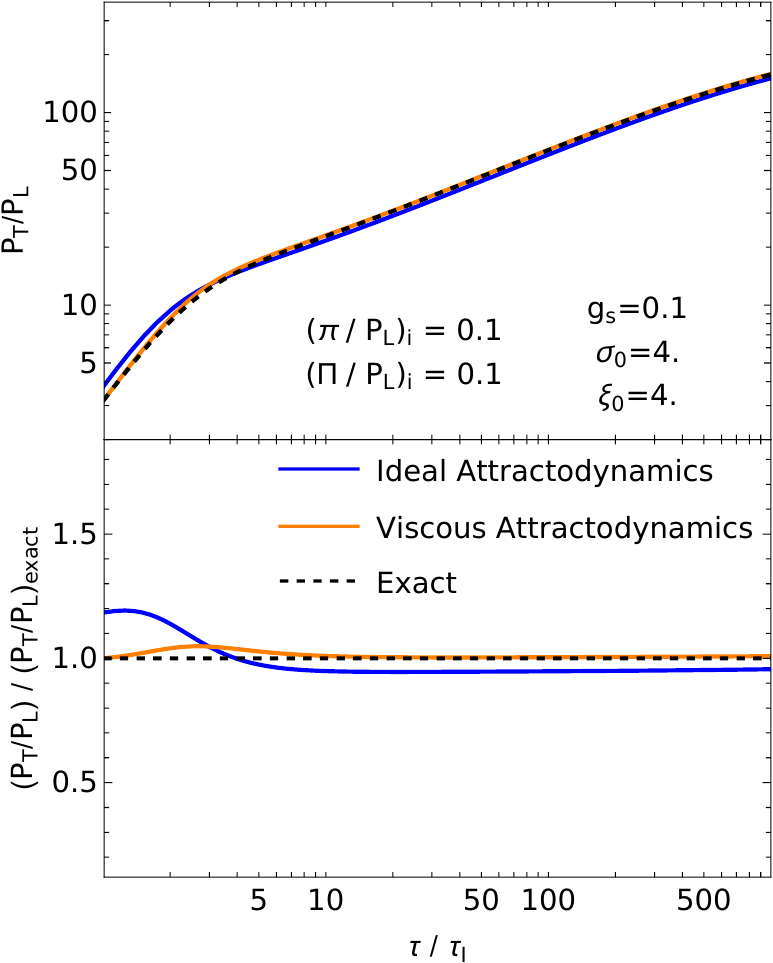}
    \includegraphics[width=0.48\linewidth]{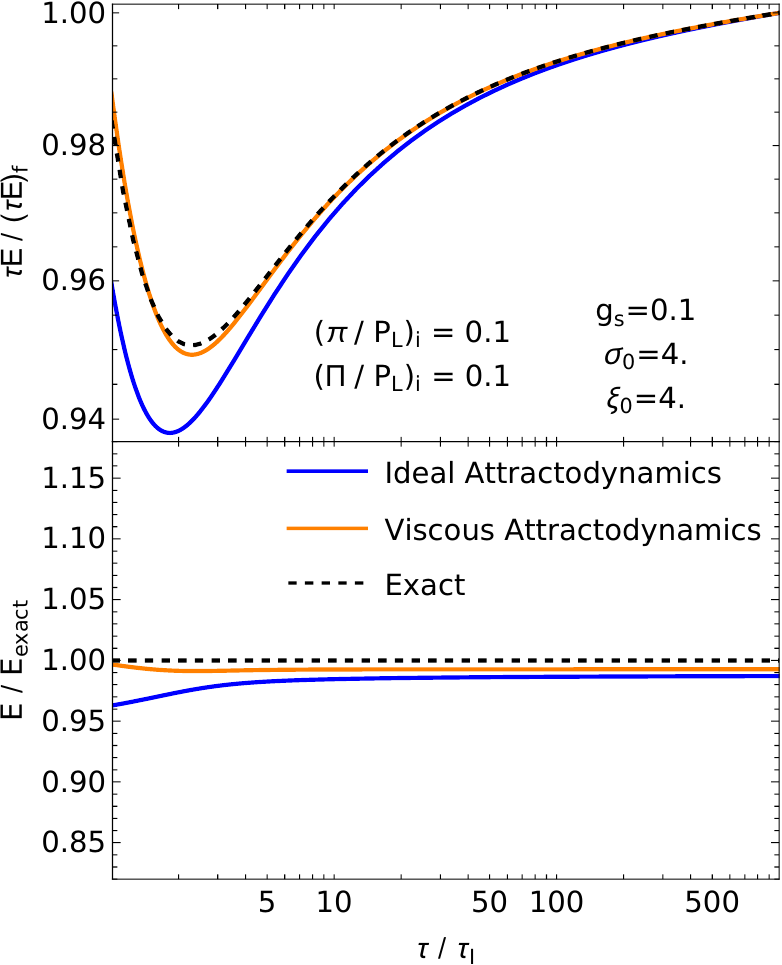}
    \caption{As in Fig.~\ref{fig:fullthy}, but here the attractodynamics curve has been derived using the linearized BSY collision kernel. The PDE solution still reflects the full BSY collision kernel.}
    \label{fig:smallpi-linear}
\end{figure}

\begin{figure}
    \centering
    \includegraphics[width=0.48\linewidth]{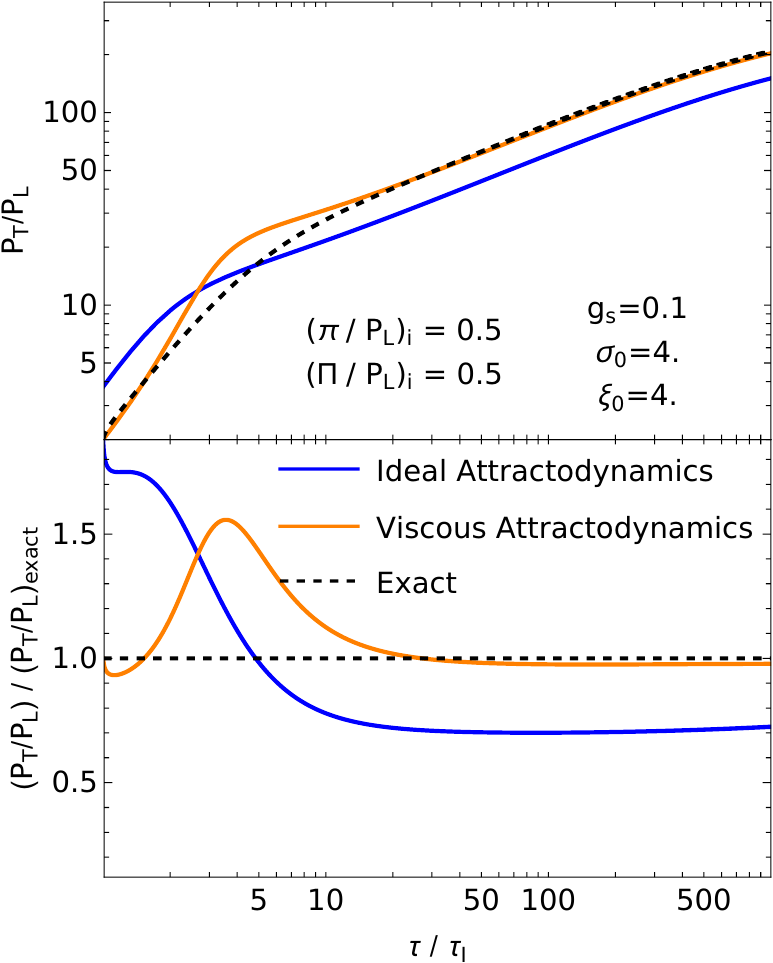}
    \includegraphics[width=0.48\linewidth]{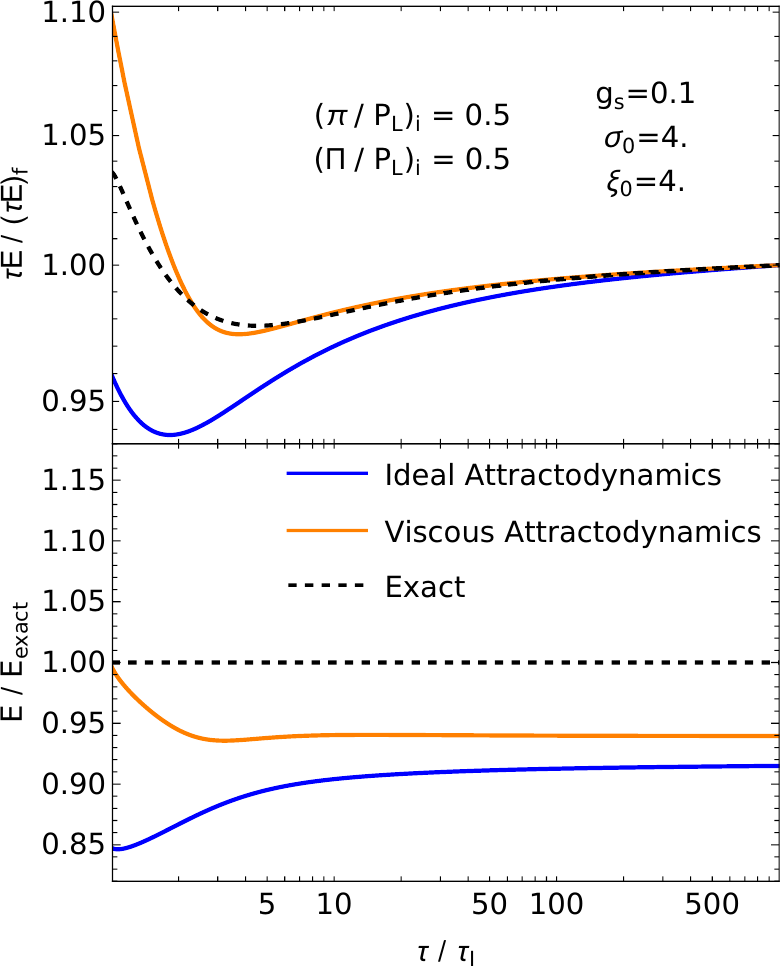}
    \caption{As in Fig.~\ref{fig:bigpi}, but here the attractodynamics curve has been derived using the linearized BSY collision kernel. The PDE solution still reflects the full BSY collision kernel.}
    \label{fig:bigpi-linear}
\end{figure}

Now we can see how well attractodynamics performs in the linearized theory by producing the equivalent of Figs.~\ref{fig:fullthy} and~\ref{fig:bigpi}. We do so in Figs.~\ref{fig:smallpi-linear} and~\ref{fig:bigpi-linear}. As expected, the performance for viscous attractodynamics derived from the linearized theory is not as close to the exact solution. However, the linearized viscous attractodynamics solution performs better than ideal attractodynamics.

\section{Scan of Initial Conditions}
\label{app:scan-initial-conditions}

\begin{figure}
    \centering
    \includegraphics[width=\linewidth]{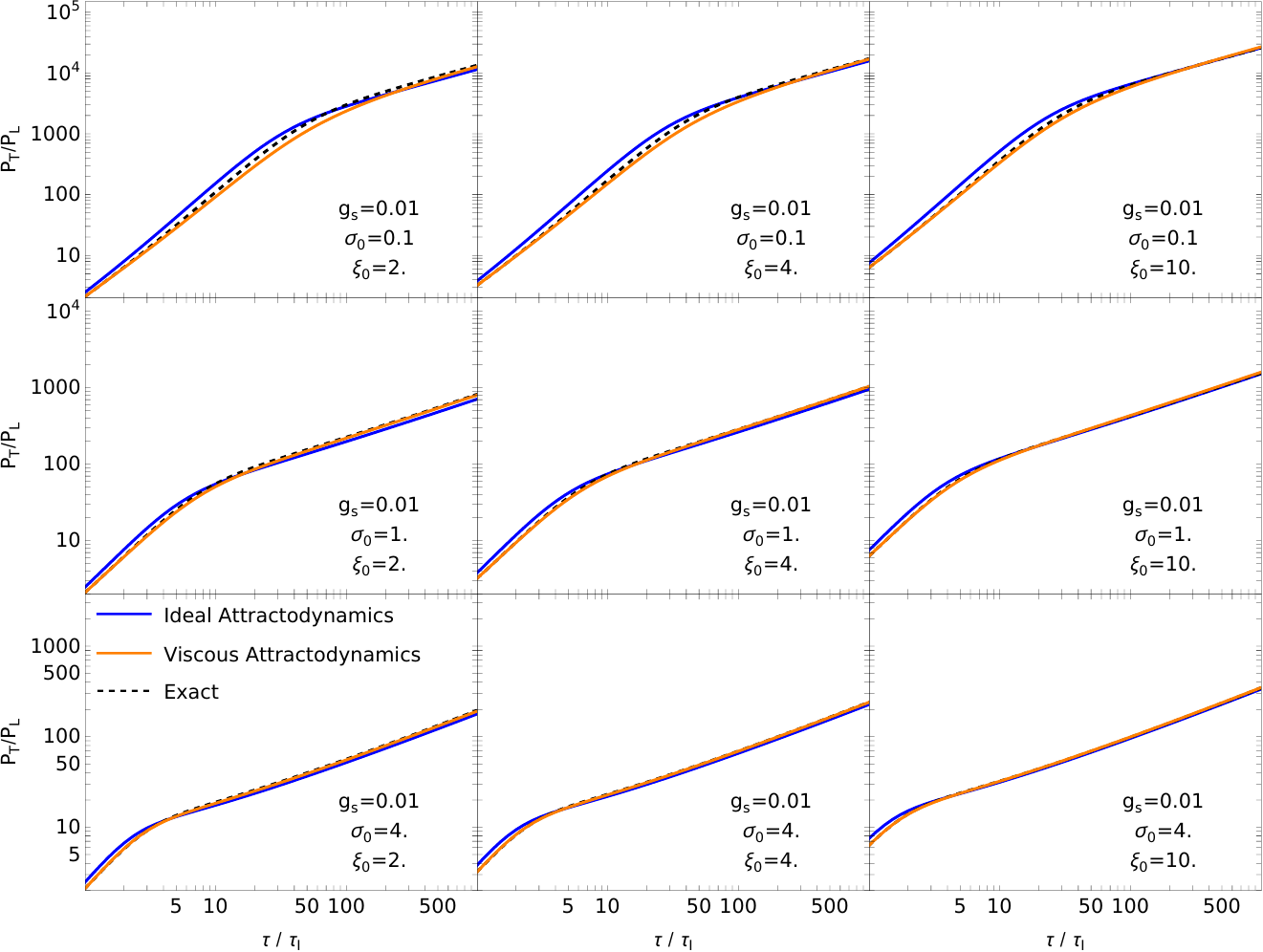}
    \caption{For $g_s=0.01$, pressure anisotropy in ideal and viscous attractodynamics compared to a full numerical solution of the Boltzmann equation for various initial occupancies and anisotropies. }
    \label{fig:scan-ic-weak}
\end{figure}

\begin{figure}
    \centering
    \includegraphics[width=\linewidth]{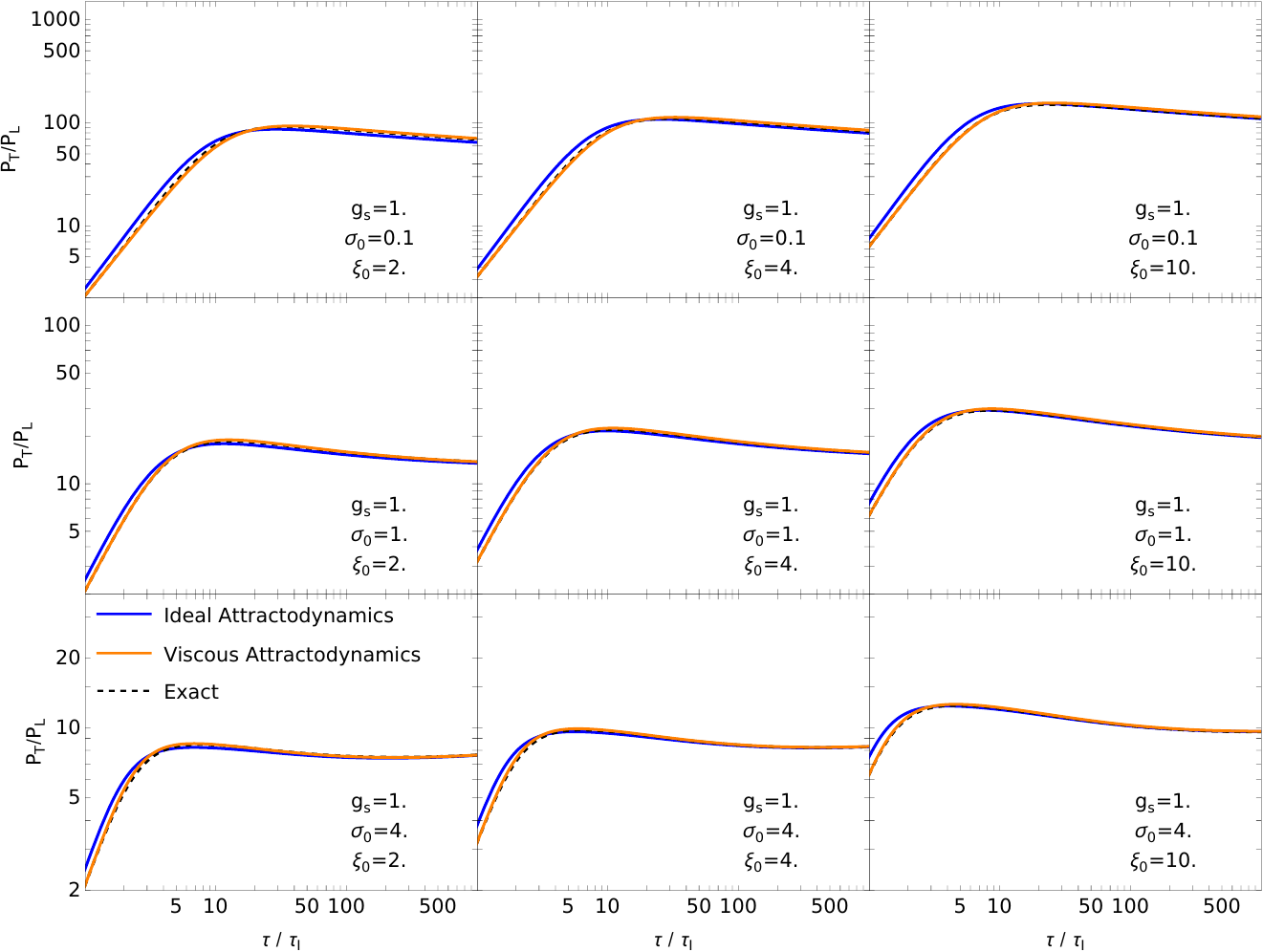}
    \caption{For $g_s=1$, pressure anisotropy in ideal and viscous attractodynamics compared to a full numerical solution of the Boltzmann equation for various initial occupancies and anisotropies.}
    \label{fig:scan-ic-strong}
\end{figure}

\begin{figure}
    \centering
    \includegraphics[width=\linewidth]{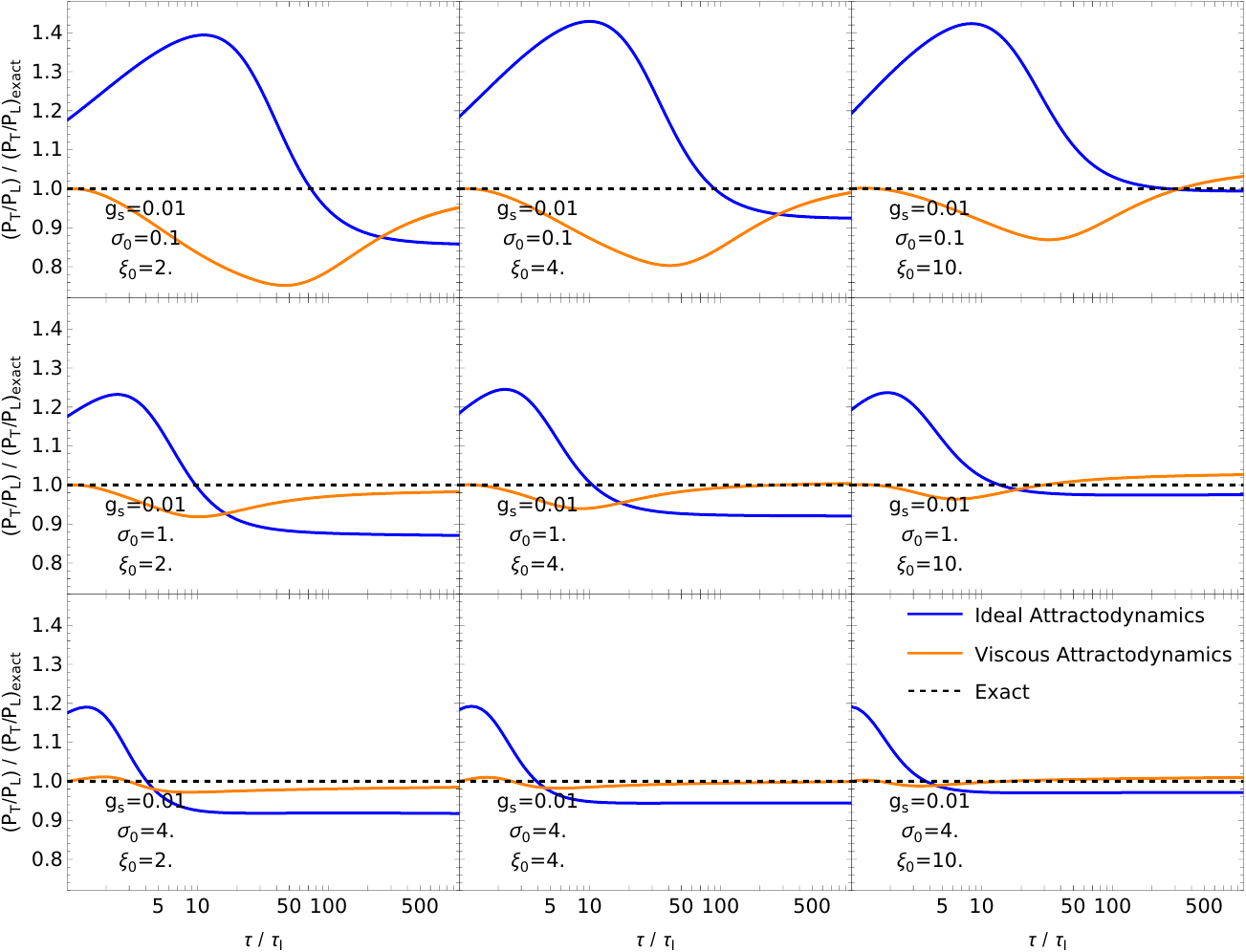}
    \caption{For $g_s=0.01$, the ratio of pressure anisotropy in ideal and viscous attractodynamics to the pressure anisotropy of a full numerical solution to the Boltzmann equation for various initial occupancies and anisotropies (as in Fig.~\ref{fig:scan-ic-weak}).}
    \label{fig:scan-ic-weak-ratio}
\end{figure}

\begin{figure}
    \centering
    \includegraphics[width=\linewidth]{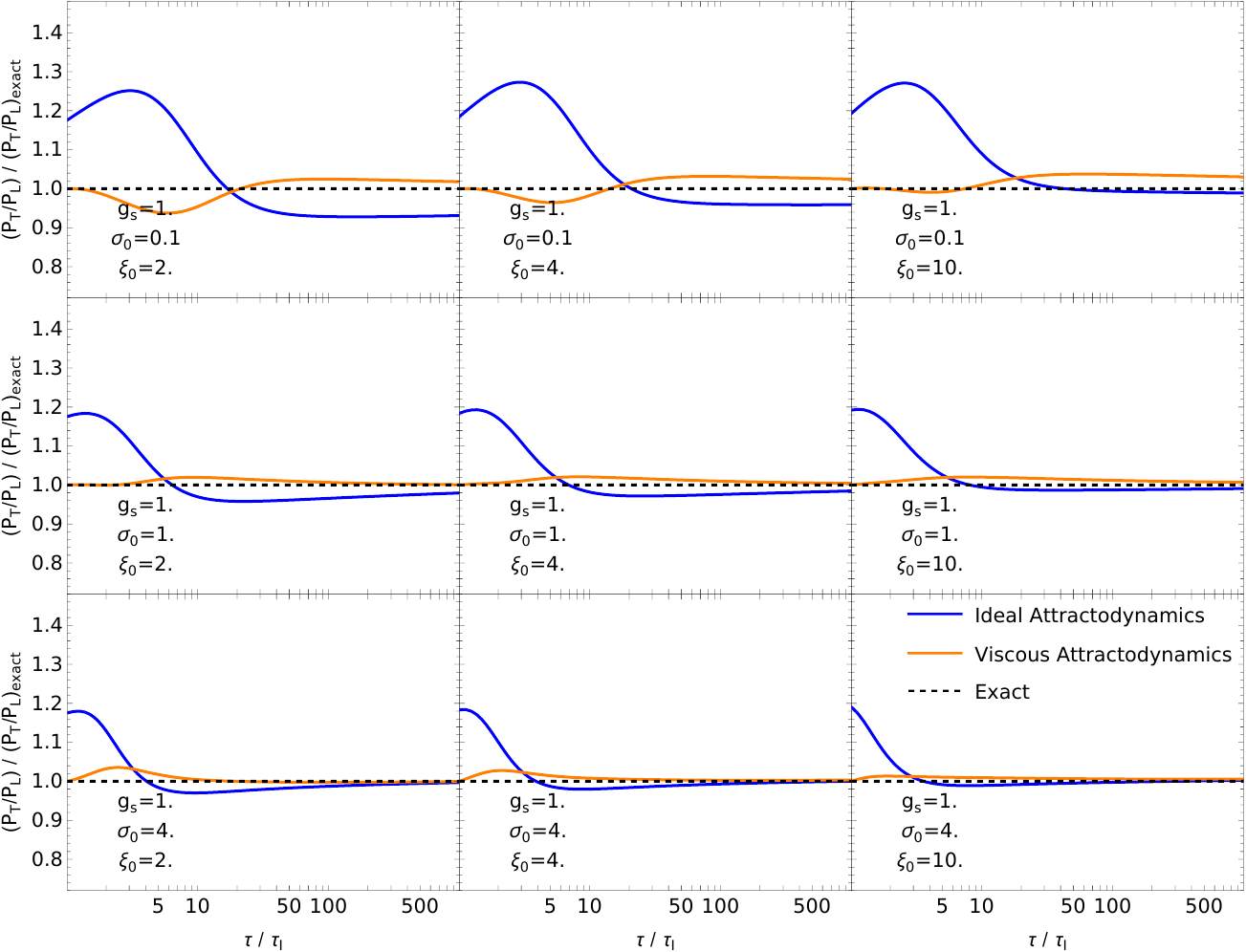}
    \caption{For $g_s=1$, the ratio of pressure anisotropy in ideal and viscous attractodynamics to the pressure anisotropy of a full numerical solution to the Boltzmann equation for various initial occupancies and anisotropies (as in Fig.~\ref{fig:scan-ic-strong}).}
    \label{fig:scan-ic-strong-ratio}
\end{figure}

\begin{figure}
    \centering
    \includegraphics[width=.9\linewidth]{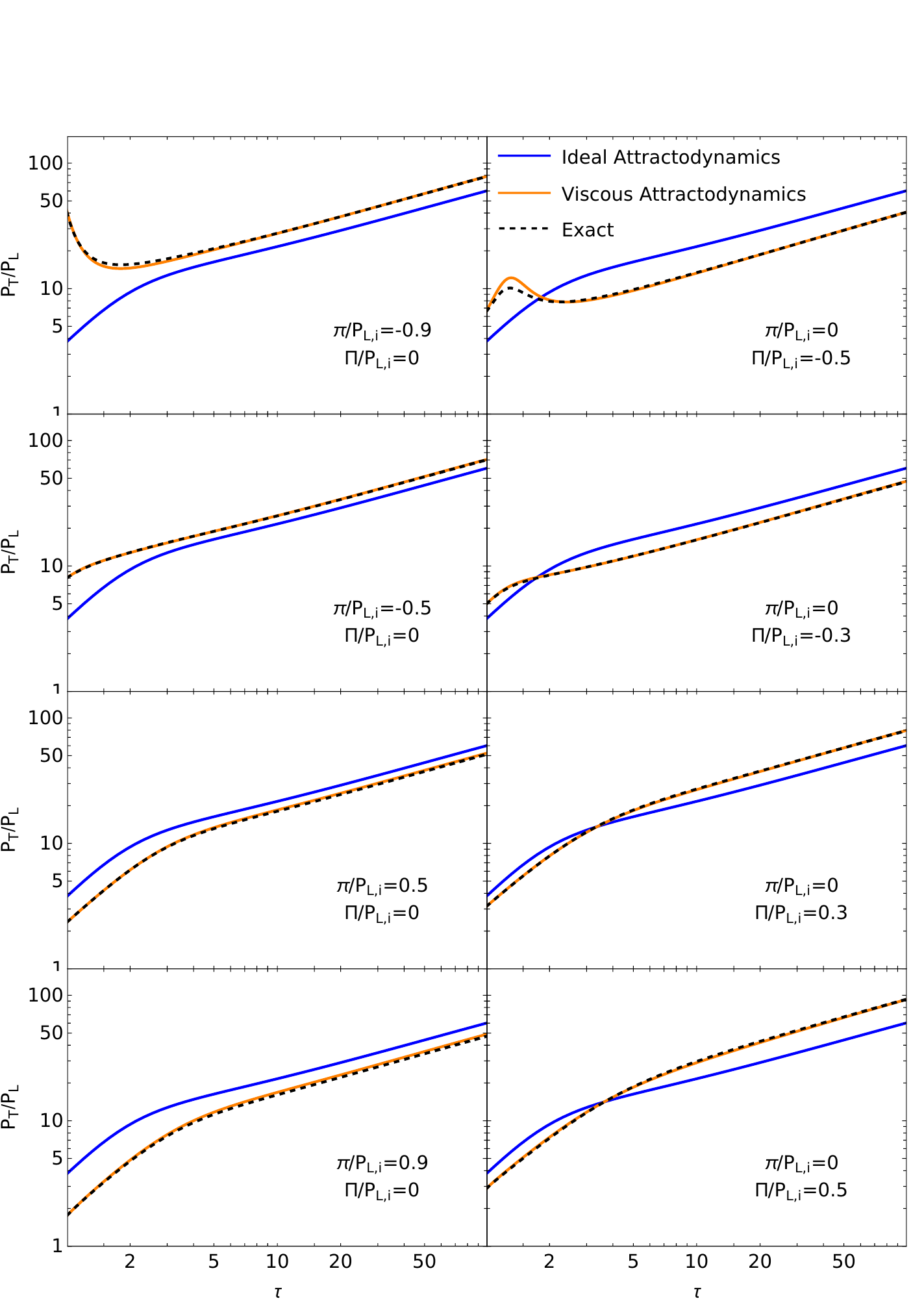}
    \caption{For $\sigma_0=4,\xi_0=4$ and a coupling $g_s=0.1$, pressure anisotropy in ideal and viscous attractodynamics compared to the pressure anisotropy of a full numerical solution to the Boltzmann equation for selected initial generalized viscous pressures $\pi$ and $\Pi$.}
    \label{fig:scan-pis}
\end{figure}

We examined one coupling, one initial condition for ideal attractodynamic variables, and two initial conditions for viscous attractodynamic variables in Sec.~\ref{sec:numerics}. To dispel any concern that these initial conditions are anomalous, here we demonstrate the broad applicability of attractodynamics by examining a broad set of possible couplings and initial conditions. In Figs.~\ref{fig:scan-ic-weak} and~\ref{fig:scan-ic-strong}, we examine pressure anisotropy in a scan over values of $\sigma_0$ and $\xi_0$ at relatively weak and strong coupling respectively for the choice  $\pi_0/(\mathcal{P}_L)_i = \Pi_0/(\mathcal{P}_L)_i = 0.1$. The pressure anisotropy for these initial conditions as a ratio to the exact solution is shown in Figs.~\ref{fig:scan-ic-weak-ratio} and~\ref{fig:scan-ic-strong-ratio}. In Fig.~\ref{fig:scan-pis}, we scan over possible initial values of $\pi$ and $\Pi$ as in Fig.~\ref{fig:attracts}, but here with comparison to the exact solution. We see that as anticipated, the attractodynamic description breaks down as the deviation from the attractor surface becomes large.

\section{Energy Non-Conservation in BSY}\label{app:energy-non-conservation}

The BSY collision kernel is an energy non-conserving approximation to the QCD EKT elastic scattering kernel in the small-angle scattering limit. If this approximation is a good one in the limit of large anisotropy, as claimed by BSY~\cite{Brewer:2022vkq}, one would expect the violation of energy conservation to be small. We can examine the degree to which this is true in practice by comparing the evolution of energy in this theory to what would be expected for Bjorken expansion in an energy conserving theory. In an energy-conserving theory, we would have
\begin{equation}
    \partial_\tau \mathcal{E} = - \frac{\mathcal{E}+\mathcal{P}_L}{\tau}\,,
\end{equation}
therefore, the dimensionless quantity
\begin{equation}
    \frac{\tau\partial_\tau \mathcal{E}}{\mathcal{E}+\mathcal{P}_L}+1
\end{equation}
characterizes the severity of energy non-conservation.
For Bjorken expansion with the BSY collision kernel, this degree of energy non-conservation has the analytic form
\begin{equation}
    \frac{\tau\partial_\tau \mathcal{E}}{\mathcal{E}+\mathcal{P}_L}+1=\frac{q m_D^2}{2g_s^2 N_c (\mathcal{E}+\mathcal{P}_L)}\,.
\end{equation}
In the viscous case we will include the contributions of $\Pi$ and $\pi$ to the energy density and longitudinal pressure, that is, we calculate
\begin{equation}
    \frac{\tau\partial_\tau (\mathcal{E}_\mathcal{A}+3\Pi)}{\mathcal{E}_\mathcal{A}+3\Pi+\mathcal{P}_{L,\mathcal{A}}+\pi+\Pi}+1\,.
\end{equation}

\begin{figure}
    \centering
    \includegraphics[width=\linewidth]{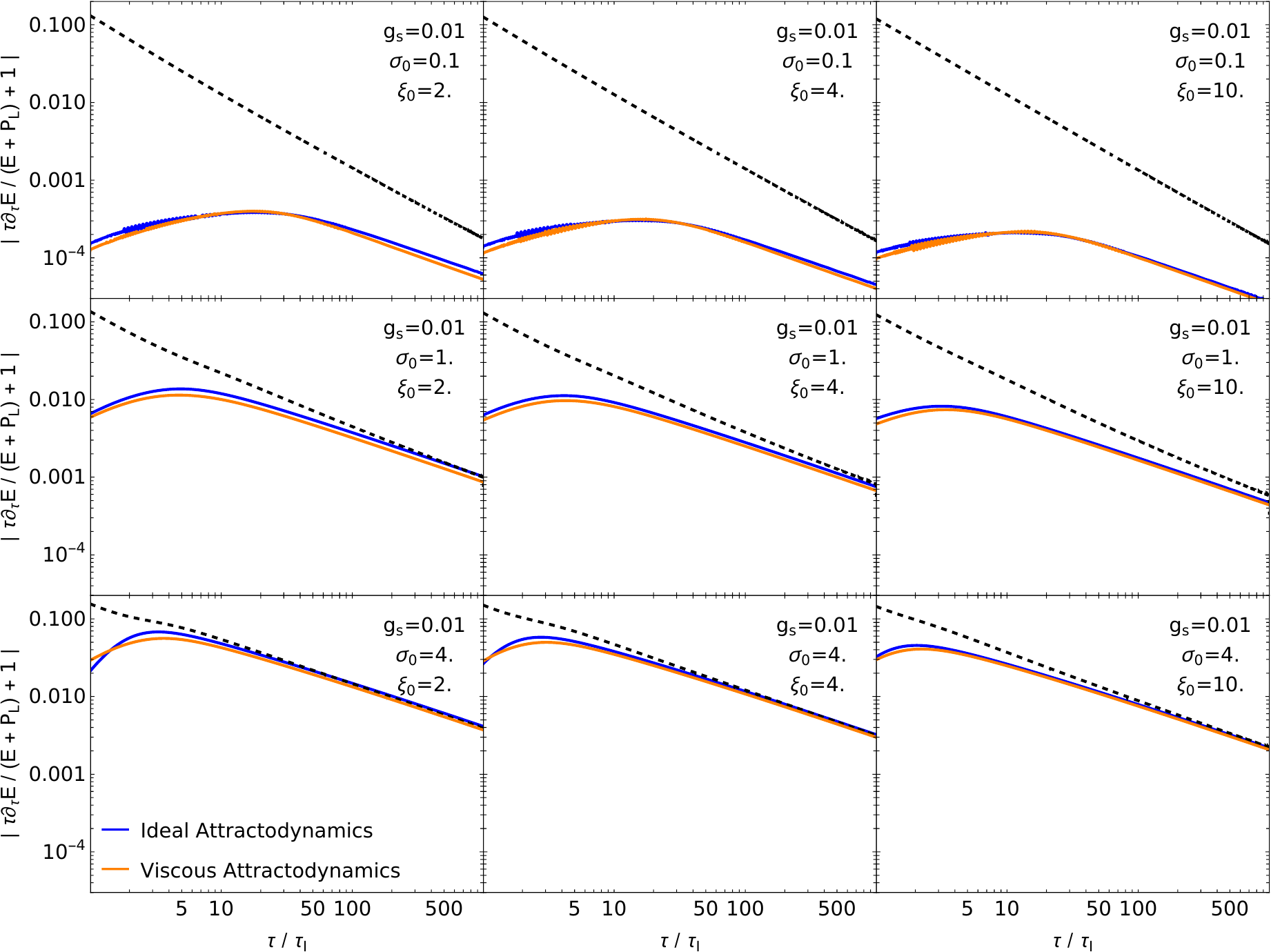}
    \caption{The degree to which energy conservation is violated in the BSY collision kernel for the same initial conditions which were considered in Fig.~\ref{fig:scan-ic-weak}.}
    \label{fig:energy-conservation-weak}
\end{figure}

\begin{figure}
    \centering
    \includegraphics[width=\linewidth]{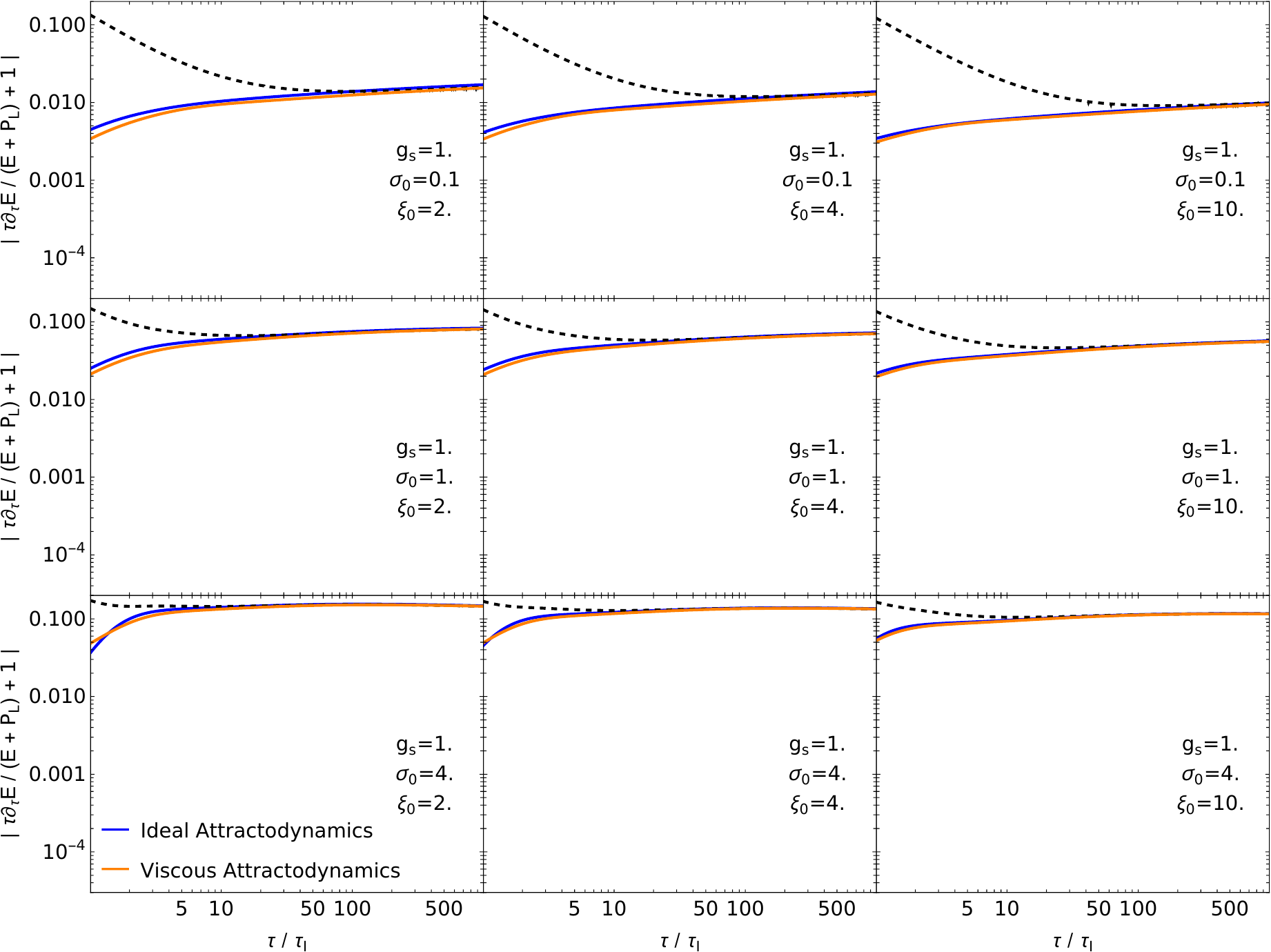}
    \caption{The degree to which energy conservation is violated in the BSY collision kernel for the same initial conditions which were considered in Fig.~\ref{fig:scan-ic-strong}.}
    \label{fig:energy-conservation-strong}
\end{figure}

For the same initial conditions analyzed in Appendix~\ref{app:scan-initial-conditions}, we plot this quantity in Figs.~\ref{fig:energy-conservation-weak} and~\ref{fig:energy-conservation-strong}. Indeed, energy is approximately conserved, with the size of deviations from energy conservation varying with initial condition.

\bibliography{draft.bib}

@article{Grad:1949zza,
    author = "Grad, Harold",
    title = "{On the kinetic theory of rarefied gases}",
    doi = "10.1002/cpa.3160020403",
    journal = "Commun. Pure Appl. Math.",
    volume = "2",
    number = "4",
    pages = "331--407",
    year = "1949"
}

@article{Berges:2025ccd,
    author = {Berges, J{\"u}rgen and Denicol, Gabriel S. and Heller, Michal P. and Preis, Thimo},
    title = "{Far from equilibrium hydrodynamics of nonthermal fixed points}",
    eprint = "2504.18754",
    archivePrefix = "arXiv",
    primaryClass = "hep-th",
    month = "4",
    year = "2025"
}

@article{Denicol:2021wod,
    author = "Denicol, Gabriel S. and Noronha, Jorge",
    editor = "Liu, Feng and Wang, Enke and Wang, Xin-Nian and Xu, Nu and Zhang, Ben-Wei",
    title = "{Fluid dynamics far-from-equilibrium: a concrete example}",
    doi = "10.1016/j.nuclphysa.2020.121996",
    journal = "Nucl. Phys. A",
    volume = "1005",
    pages = "121996",
    year = "2021"
}

@article{Romatschke:2003ms,
    author = "Romatschke, Paul and Strickland, Michael",
    title = "{Collective modes of an anisotropic quark gluon plasma}",
    eprint = "hep-ph/0304092",
    archivePrefix = "arXiv",
    reportNumber = "TUW-03-09",
    doi = "10.1103/PhysRevD.68.036004",
    journal = "Phys. Rev. D",
    volume = "68",
    pages = "036004",
    year = "2003"
}

@article{Glidden:2020qmu,
    author = "Glidden, Jake A. P. and Eigen, Christoph and Dogra, Lena H. and Hilker, Timon A. and Smith, Robert P. and Hadzibabic, Zoran",
    title = "{Bidirectional dynamic scaling in an isolated Bose gas far from equilibrium}",
    eprint = "2006.01118",
    archivePrefix = "arXiv",
    primaryClass = "cond-mat.quant-gas",
    doi = "10.1038/s41567-020-01114-x",
    journal = "Nature Phys.",
    volume = "17",
    number = "4",
    pages = "457--461",
    year = "2021"
}

@article{Prufer:2018hto,
    author = {Pr{\"u}fer, Maximilian and Kunkel, Philipp and Strobel, Helmut and Lannig, Stefan and Linnemann, Daniel and Schmied, Christian-Marcel and Berges, J{\"u}rgen and Gasenzer, Thomas and Oberthaler, Markus K.},
    title = "{Observation of universal dynamics in a spinor Bose gas far from equilibrium}",
    eprint = "1805.11881",
    archivePrefix = "arXiv",
    primaryClass = "cond-mat.quant-gas",
    doi = "10.1038/s41586-018-0659-0",
    journal = "Nature",
    volume = "563",
    number = "7730",
    pages = "217--220",
    year = "2018"
}

@article{Berges:2004yj,
    author = "Berges, Juergen",
    editor = "Bracco, Mirian and Chiapparini, Marcelo and Ferreira, Erasmo and Kodama, Takeshi",
    title = "{Introduction to nonequilibrium quantum field theory}",
    eprint = "hep-ph/0409233",
    archivePrefix = "arXiv",
    doi = "10.1063/1.1843591",
    journal = "AIP Conf. Proc.",
    volume = "739",
    number = "1",
    pages = "3--62",
    year = "2004"
}

@article{Berges:2015kfa,
    author = "Berges, Jurgen",
    title = "{Nonequilibrium Quantum Fields: From Cold Atoms to Cosmology}",
    eprint = "1503.02907",
    archivePrefix = "arXiv",
    primaryClass = "hep-ph",
    month = "3",
    year = "2015"
}

@article{Heller:2016rtz,
    author = "Heller, Michal P. and Kurkela, Aleksi and Spali{\'n}ski, Michal and Svensson, Viktor",
    title = "{Hydrodynamization in kinetic theory: Transient modes and the gradient expansion}",
    eprint = "1609.04803",
    archivePrefix = "arXiv",
    primaryClass = "nucl-th",
    reportNumber = "CERN-TH-2016-199",
    doi = "10.1103/PhysRevD.97.091503",
    journal = "Phys. Rev. D",
    volume = "97",
    number = "9",
    pages = "091503",
    year = "2018"
}

@article{Mitra:2020mei,
    author = "Mitra, Toshali and Mondkar, Sukrut and Mukhopadhyay, Ayan and Rebhan, Anton and Soloviev, Alexander",
    title = "{Hydrodynamic attractor of a hybrid viscous fluid in Bjorken flow}",
    eprint = "2006.09383",
    archivePrefix = "arXiv",
    primaryClass = "hep-ph",
    doi = "10.1103/PhysRevResearch.2.043320",
    journal = "Phys. Rev. Res.",
    volume = "2",
    number = "4",
    pages = "043320",
    year = "2020"
}

@article{Kurkela:2018wud,
    author = {Kurkela, Aleksi and Mazeliauskas, Aleksas and Paquet, Jean-Fran{\c{c}}ois and Schlichting, S{\"o}ren and Teaney, Derek},
    title = "{Matching the Nonequilibrium Initial Stage of Heavy Ion Collisions to Hydrodynamics with QCD Kinetic Theory}",
    eprint = "1805.01604",
    archivePrefix = "arXiv",
    primaryClass = "hep-ph",
    doi = "10.1103/PhysRevLett.122.122302",
    journal = "Phys. Rev. Lett.",
    volume = "122",
    number = "12",
    pages = "122302",
    year = "2019"
}

@article{Heller:2011ju,
    author = "Heller, Michal P. and Janik, Romuald A. and Witaszczyk, Przemyslaw",
    title = "{The characteristics of thermalization of boost-invariant plasma from holography}",
    eprint = "1103.3452",
    archivePrefix = "arXiv",
    primaryClass = "hep-th",
    doi = "10.1103/PhysRevLett.108.201602",
    journal = "Phys. Rev. Lett.",
    volume = "108",
    pages = "201602",
    year = "2012"
}

@article{Baier:2000sb,
    author = "Baier, R. and Mueller, Alfred H. and Schiff, D. and Son, D. T.",
    title = "{'Bottom up' thermalization in heavy ion collisions}",
    eprint = "hep-ph/0009237",
    archivePrefix = "arXiv",
    doi = "10.1016/S0370-2693(01)00191-5",
    journal = "Phys. Lett. B",
    volume = "502",
    pages = "51--58",
    year = "2001"
}

@article{Israel:1979wp,
    author = "Israel, W. and Stewart, J. M.",
    title = "{Transient relativistic thermodynamics and kinetic theory}",
    doi = "10.1016/0003-4916(79)90130-1",
    journal = "Annals Phys.",
    volume = "118",
    pages = "341--372",
    year = "1979"
}

@article{Alqahtani:2017mhy,
    author = "Alqahtani, Mubarak and Nopoush, Mohammad and Strickland, Michael",
    title = "{Relativistic anisotropic hydrodynamics}",
    eprint = "1712.03282",
    archivePrefix = "arXiv",
    primaryClass = "nucl-th",
    doi = "10.1016/j.ppnp.2018.05.004",
    journal = "Prog. Part. Nucl. Phys.",
    volume = "101",
    pages = "204--248",
    year = "2018"
}

@article{Florkowski:2016zsi,
    author = "Florkowski, Wojciech and Ryblewski, Radoslaw and Spali{\'n}ski, Micha{\l}",
    title = "{Gradient expansion for anisotropic hydrodynamics}",
    eprint = "1608.07558",
    archivePrefix = "arXiv",
    primaryClass = "nucl-th",
    doi = "10.1103/PhysRevD.94.114025",
    journal = "Phys. Rev. D",
    volume = "94",
    number = "11",
    pages = "114025",
    year = "2016"
}

@article{Kovtun:2012rj,
    author = "Kovtun, Pavel",
    title = "{Lectures on hydrodynamic fluctuations in relativistic theories}",
    eprint = "1205.5040",
    archivePrefix = "arXiv",
    primaryClass = "hep-th",
    doi = "10.1088/1751-8113/45/47/473001",
    journal = "J. Phys. A",
    volume = "45",
    pages = "473001",
    year = "2012"
}

@article{Crossley:2015evo,
    author = "Crossley, Michael and Glorioso, Paolo and Liu, Hong",
    title = "{Effective field theory of dissipative fluids}",
    eprint = "1511.03646",
    archivePrefix = "arXiv",
    primaryClass = "hep-th",
    reportNumber = "MIT-CTP-4734",
    doi = "10.1007/JHEP09(2017)095",
    journal = "JHEP",
    volume = "09",
    pages = "095",
    year = "2017"
}

@article{Liu:2018kfw,
    author = "Liu, Hong and Glorioso, Paolo",
    title = "{Lectures on non-equilibrium effective field theories and fluctuating hydrodynamics}",
    eprint = "1805.09331",
    archivePrefix = "arXiv",
    primaryClass = "hep-th",
    reportNumber = "MIT-CTP/5018; EFI-18-8, MIT-CTP-5018, EFI-18-8",
    doi = "10.22323/1.305.0008",
    journal = "PoS",
    volume = "TASI2017",
    pages = "008",
    year = "2018"
}

@article{Romatschke:2016hle,
    author = "Romatschke, Paul",
    title = "{Do nuclear collisions create a locally equilibrated quark{\textendash}gluon plasma?}",
    eprint = "1609.02820",
    archivePrefix = "arXiv",
    primaryClass = "nucl-th",
    doi = "10.1140/epjc/s10052-016-4567-x",
    journal = "Eur. Phys. J. C",
    volume = "77",
    number = "1",
    pages = "21",
    year = "2017"
}

@article{Romatschke:2017vte,
    author = "Romatschke, Paul",
    title = "{Relativistic Fluid Dynamics Far From Local Equilibrium}",
    eprint = "1704.08699",
    archivePrefix = "arXiv",
    primaryClass = "hep-th",
    doi = "10.1103/PhysRevLett.120.012301",
    journal = "Phys. Rev. Lett.",
    volume = "120",
    number = "1",
    pages = "012301",
    year = "2018"
}

@article{Arnold:2002zm,
    author = "Arnold, Peter Brockway and Moore, Guy D. and Yaffe, Laurence G.",
    title = "{Effective kinetic theory for high temperature gauge theories}",
    eprint = "hep-ph/0209353",
    archivePrefix = "arXiv",
    doi = "10.1088/1126-6708/2003/01/030",
    journal = "JHEP",
    volume = "01",
    pages = "030",
    year = "2003"
}

@article{Martinez:2010sc,
    author = "Martinez, Mauricio and Strickland, Michael",
    title = "{Dissipative Dynamics of Highly Anisotropic Systems}",
    eprint = "1007.0889",
    archivePrefix = "arXiv",
    primaryClass = "nucl-th",
    doi = "10.1016/j.nuclphysa.2010.08.011",
    journal = "Nucl. Phys. A",
    volume = "848",
    pages = "183--197",
    year = "2010"
}

@article{Florkowski:2010cf,
    author = "Florkowski, Wojciech and Ryblewski, Radoslaw",
    title = "{Highly-anisotropic and strongly-dissipative hydrodynamics for early stages of relativistic heavy-ion collisions}",
    eprint = "1007.0130",
    archivePrefix = "arXiv",
    primaryClass = "nucl-th",
    doi = "10.1103/PhysRevC.83.034907",
    journal = "Phys. Rev. C",
    volume = "83",
    pages = "034907",
    year = "2011"
}

@article{Bazow:2013ifa,
    author = "Bazow, Dennis and Heinz, Ulrich W. and Strickland, Michael",
    title = "{Second-order (2+1)-dimensional anisotropic hydrodynamics}",
    eprint = "1311.6720",
    archivePrefix = "arXiv",
    primaryClass = "nucl-th",
    doi = "10.1103/PhysRevC.90.054910",
    journal = "Phys. Rev. C",
    volume = "90",
    number = "5",
    pages = "054910",
    year = "2014"
}

@article{Mueller:1999pi,
    author = "Mueller, Alfred H.",
    title = "{The Boltzmann equation for gluons at early times after a heavy ion collision}",
    eprint = "hep-ph/9909388",
    archivePrefix = "arXiv",
    reportNumber = "CU-TP-950",
    doi = "10.1016/S0370-2693(00)00084-8",
    journal = "Phys. Lett. B",
    volume = "475",
    pages = "220--224",
    year = "2000"
}

@article{Brewer:2022vkq,
    author = "Brewer, Jasmine and Scheihing-Hitschfeld, Bruno and Yin, Yi",
    title = "{Scaling and adiabaticity in a rapidly expanding gluon plasma}",
    eprint = "2203.02427",
    archivePrefix = "arXiv",
    primaryClass = "hep-ph",
    reportNumber = "CERN-TH-2022-026, MIT-CTP/5411",
    doi = "10.1007/JHEP05(2022)145",
    journal = "JHEP",
    volume = "05",
    pages = "145",
    year = "2022"
}

@article{Brewer:2019oha,
    author = "Brewer, Jasmine and Yan, Li and Yin, Yi",
    title = "{Adiabatic hydrodynamization in rapidly-expanding quark{\textendash}gluon plasma}",
    eprint = "1910.00021",
    archivePrefix = "arXiv",
    primaryClass = "nucl-th",
    reportNumber = "MIT-CTP/5141",
    doi = "10.1016/j.physletb.2021.136189",
    journal = "Phys. Lett. B",
    volume = "816",
    pages = "136189",
    year = "2021"
}

@article{Rajagopal:2024lou,
    author = "Rajagopal, Krishna and Scheihing-Hitschfeld, Bruno and Steinhorst, Rachel",
    title = "{Adiabatic Hydrodynamization and the emergence of attractors: a unified description of hydrodynamization in kinetic theory}",
    eprint = "2405.17545",
    archivePrefix = "arXiv",
    primaryClass = "hep-ph",
    reportNumber = "MIT-CTP/5724",
    doi = "10.1007/JHEP04(2025)028",
    journal = "JHEP",
    volume = "04",
    pages = "028",
    year = "2025"
}

@article{Gorghetto:2018myk,
    author = "Gorghetto, Marco and Hardy, Edward and Villadoro, Giovanni",
    title = "{Axions from Strings: the Attractive Solution}",
    eprint = "1806.04677",
    archivePrefix = "arXiv",
    primaryClass = "hep-ph",
    doi = "10.1007/JHEP07(2018)151",
    journal = "JHEP",
    volume = "07",
    pages = "151",
    year = "2018"
}

@article{Hindmarsh:2021vih,
    author = "Hindmarsh, Mark and Lizarraga, Joanes and Lopez-Eiguren, Asier and Urrestilla, Jon",
    title = "{Approach to scaling in axion string networks}",
    eprint = "2102.07723",
    archivePrefix = "arXiv",
    primaryClass = "astro-ph.CO",
    reportNumber = "HIP-2021-7/TH",
    doi = "10.1103/PhysRevD.103.103534",
    journal = "Phys. Rev. D",
    volume = "103",
    number = "10",
    pages = "103534",
    year = "2021"
}

@article{Blaizot:2013lga,
    author = "Blaizot, Jean-Paul and Liao, Jinfeng and McLerran, Larry",
    title = "{Gluon Transport Equation in the Small Angle Approximation and the Onset of Bose-Einstein Condensation}",
    eprint = "1305.2119",
    archivePrefix = "arXiv",
    primaryClass = "hep-ph",
    doi = "10.1016/j.nuclphysa.2013.10.010",
    journal = "Nucl. Phys. A",
    volume = "920",
    pages = "58--77",
    year = "2013"
}

@article{DeLescluze:2025gaa,
    author = "De Lescluze, Matisse and Heller, Michal P. and Mazeliauskas, Aleksas and Scheihing-Hitschfeld, Bruno and Werthmann, Clemens",
    title = "{Adiabatic hydrodynamization and quasinormal modes of nonthermal attractors}",
    eprint = "2510.15016",
    archivePrefix = "arXiv",
    primaryClass = "hep-ph",
    month = "10",
    year = "2025"
}

@article{Mikheev:2022fdl,
    author = {Mikheev, Aleksandr N. and Mazeliauskas, Aleksas and Berges, J{\"u}rgen},
    title = "{Stability analysis of nonthermal fixed points in longitudinally expanding kinetic theory}",
    eprint = "2203.02299",
    archivePrefix = "arXiv",
    primaryClass = "hep-ph",
    reportNumber = "CERN-TH-2022-030",
    doi = "10.1103/PhysRevD.105.116025",
    journal = "Phys. Rev. D",
    volume = "105",
    number = "11",
    pages = "116025",
    year = "2022"
}

@article{Berges:2013fga,
    author = "Berges, Juergen and Boguslavski, Kirill and Schlichting, Soeren and Venugopalan, Raju",
    title = "{Universal attractor in a highly occupied non-Abelian plasma}",
    eprint = "1311.3005",
    archivePrefix = "arXiv",
    primaryClass = "hep-ph",
    doi = "10.1103/PhysRevD.89.114007",
    journal = "Phys. Rev. D",
    volume = "89",
    number = "11",
    pages = "114007",
    year = "2014"
}

@article{Berges:2013eia,
    author = "Berges, J. and Boguslavski, K. and Schlichting, S. and Venugopalan, R.",
    title = "{Turbulent thermalization process in heavy-ion collisions at ultrarelativistic energies}",
    eprint = "1303.5650",
    archivePrefix = "arXiv",
    primaryClass = "hep-ph",
    doi = "10.1103/PhysRevD.89.074011",
    journal = "Phys. Rev. D",
    volume = "89",
    number = "7",
    pages = "074011",
    year = "2014"
}

@article{Martinez:2012tu,
    author = "Martinez, Mauricio and Ryblewski, Radoslaw and Strickland, Michael",
    title = "{Boost-Invariant (2+1)-dimensional Anisotropic Hydrodynamics}",
    eprint = "1204.1473",
    archivePrefix = "arXiv",
    primaryClass = "nucl-th",
    reportNumber = "INT-PUB-12-014",
    doi = "10.1103/PhysRevC.85.064913",
    journal = "Phys. Rev. C",
    volume = "85",
    pages = "064913",
    year = "2012"
}

@article{Rajagopal:2025nca,
    author = "Rajagopal, Krishna and Scheihing-Hitschfeld, Bruno and Steinhorst, Rachel",
    title = "{Attractors without scaling: adiabatic hydrodynamization with and without inelastic scattering}",
    eprint = "2507.21232",
    archivePrefix = "arXiv",
    primaryClass = "hep-ph",
    reportNumber = "MIT-CTP/5898",
    doi = "10.1007/JHEP03(2026)003",
    journal = "JHEP",
    volume = "03",
    pages = "003",
    year = "2026"
}

@article{Kurkela:2012hp,
    author = "Kurkela, Aleksi and Moore, Guy D.",
    title = "{UV Cascade in Classical Yang-Mills Theory}",
    eprint = "1207.1663",
    archivePrefix = "arXiv",
    primaryClass = "hep-ph",
    reportNumber = "INT-PUB-12-032",
    doi = "10.1103/PhysRevD.86.056008",
    journal = "Phys. Rev. D",
    volume = "86",
    pages = "056008",
    year = "2012"
}

@article{Schlichting:2012es,
    author = "Schlichting, Soeren",
    title = "{Turbulent thermalization of weakly coupled non-abelian plasmas}",
    eprint = "1207.1450",
    archivePrefix = "arXiv",
    primaryClass = "hep-ph",
    doi = "10.1103/PhysRevD.86.065008",
    journal = "Phys. Rev. D",
    volume = "86",
    pages = "065008",
    year = "2012"
}

@article{Berges:2013lsa,
    author = "Berges, J. and Boguslavski, K. and Schlichting, S. and Venugopalan, R.",
    title = "{Basin of attraction for turbulent thermalization and the range of validity of classical-statistical simulations}",
    eprint = "1312.5216",
    archivePrefix = "arXiv",
    primaryClass = "hep-ph",
    doi = "10.1007/JHEP05(2014)054",
    journal = "JHEP",
    volume = "05",
    pages = "054",
    year = "2014"
}

@article{AbraaoYork:2014hbk,
    author = "Abraao York, Mark C. and Kurkela, Aleksi and Lu, Egang and Moore, Guy D.",
    title = "{UV cascade in classical Yang-Mills theory via kinetic theory}",
    eprint = "1401.3751",
    archivePrefix = "arXiv",
    primaryClass = "hep-ph",
    reportNumber = "CERN-PH-TH-2014-006",
    doi = "10.1103/PhysRevD.89.074036",
    journal = "Phys. Rev. D",
    volume = "89",
    number = "7",
    pages = "074036",
    year = "2014"
}

@article{Mikheev:2018adp,
    author = "Mikheev, Aleksandr N. and Schmied, Christian-Marcel and Gasenzer, Thomas",
    title = "{Low-energy effective theory of nonthermal fixed points in a multicomponent Bose gas}",
    eprint = "1807.10228",
    archivePrefix = "arXiv",
    primaryClass = "cond-mat.quant-gas",
    doi = "10.1103/PhysRevA.99.063622",
    journal = "Phys. Rev. A",
    volume = "99",
    number = "6",
    pages = "063622",
    year = "2019"
}

@article{PineiroOrioli:2015cpb,
    author = "Pi{\~n}eiro Orioli, A. and Boguslavski, K. and Berges, J.",
    title = "{Universal self-similar dynamics of relativistic and nonrelativistic field theories near nonthermal fixed points}",
    eprint = "1503.02498",
    archivePrefix = "arXiv",
    primaryClass = "hep-ph",
    doi = "10.1103/PhysRevD.92.025041",
    journal = "Phys. Rev. D",
    volume = "92",
    number = "2",
    pages = "025041",
    year = "2015"
}

@article{Schlichting:2019abc,
    author = "Schlichting, Soeren and Teaney, Derek",
    title = "{The First fm/c of Heavy-Ion Collisions}",
    eprint = "1908.02113",
    archivePrefix = "arXiv",
    primaryClass = "nucl-th",
    doi = "10.1146/annurev-nucl-101918-023825",
    journal = "Ann. Rev. Nucl. Part. Sci.",
    volume = "69",
    pages = "447--476",
    year = "2019"
}

@article{Berges:2020fwq,
    author = {Berges, J{\"u}rgen and Heller, Michal P. and Mazeliauskas, Aleksas and Venugopalan, Raju},
    title = "{QCD thermalization: Ab initio approaches and interdisciplinary connections}",
    eprint = "2005.12299",
    archivePrefix = "arXiv",
    primaryClass = "hep-th",
    reportNumber = "CERN-TH-2020-080",
    doi = "10.1103/RevModPhys.93.035003",
    journal = "Rev. Mod. Phys.",
    volume = "93",
    number = "3",
    pages = "035003",
    year = "2021"
}

@article{Erne:2018gmz,
    author = {Erne, Sebastian and B{\"u}cker, Robert and Gasenzer, Thomas and Berges, J{\"u}rgen and Schmiedmayer, J{\"o}rg},
    title = "{Universal dynamics in an isolated one-dimensional Bose gas far from equilibrium}",
    eprint = "1805.12310",
    archivePrefix = "arXiv",
    primaryClass = "cond-mat.quant-gas",
    doi = "10.1038/s41586-018-0667-0",
    journal = "Nature",
    volume = "563",
    number = "7730",
    pages = "225--229",
    year = "2018"
}

@article{Huh:2023xso,
    author = "Huh, SeungJung and Mukherjee, Koushik and Kwon, Kiryang and Seo, Jihoon and Hur, Junhyeok and Mistakidis, Simeon I. and Sadeghpour, H. R. and Choi, Jae-yoon",
    title = "{Universality class of a spinor Bose{\textendash}Einstein condensate far from equilibrium}",
    eprint = "2303.05230",
    archivePrefix = "arXiv",
    primaryClass = "cond-mat.quant-gas",
    doi = "10.1038/s41567-023-02339-2",
    journal = "Nature Phys.",
    volume = "20",
    number = "3",
    pages = "402--408",
    year = "2024"
}

@article{Lannig:2023fzf,
    author = {Lannig, Stefan and Pr{\"u}fer, Maximilian and Deller, Yannick and Siovitz, Ido and Dreher, Jan and Gasenzer, Thomas and Strobel, Helmut and Oberthaler, Markus K.},
    title = "{Observation of two non-thermal fixed points for the same microscopic symmetry}",
    eprint = "2306.16497",
    archivePrefix = "arXiv",
    primaryClass = "cond-mat.quant-gas",
    month = "6",
    year = "2023"
}

@article{Martirosyan:2023mml,
    author = "Martirosyan, Gevorg and Ho, Christopher J. and Etrych, Ji{\v{r}}{\'\i} and Zhang, Yansheng and Cao, Alec and Hadzibabic, Zoran and Eigen, Christoph",
    title = "{Observation of Subdiffusive Dynamic Scaling in a Driven and Disordered Bose Gas}",
    eprint = "2304.06697",
    archivePrefix = "arXiv",
    primaryClass = "cond-mat.quant-gas",
    doi = "10.1103/PhysRevLett.132.113401",
    journal = "Phys. Rev. Lett.",
    volume = "132",
    number = "11",
    pages = "113401",
    year = "2024"
}

@article{Gazo:2023exc,
    author = "Gazo, Martin and Karailiev, Andrey and Satoor, Tanish and Eigen, Christoph and Ga{\l}ka, Maciej and Hadzibabic, Zoran",
    title = "{Universal coarsening in a homogeneous two-dimensional Bose gas}",
    eprint = "2312.09248",
    archivePrefix = "arXiv",
    primaryClass = "cond-mat.quant-gas",
    doi = "10.1126/science.ado3487",
    journal = "Science",
    volume = "389",
    number = "6762",
    pages = "ado3487",
    year = "2025"
}

@article{Kurkela:2015qoa,
    author = "Kurkela, Aleksi and Zhu, Yan",
    title = "{Isotropization and hydrodynamization in weakly coupled heavy-ion collisions}",
    eprint = "1506.06647",
    archivePrefix = "arXiv",
    primaryClass = "hep-ph",
    reportNumber = "CERN-PH-TH-2015-142",
    doi = "10.1103/PhysRevLett.115.182301",
    journal = "Phys. Rev. Lett.",
    volume = "115",
    number = "18",
    pages = "182301",
    year = "2015"
}

@article{Micha:2002ey,
    author = "Micha, Raphael and Tkachev, Igor I.",
    title = "{Relativistic turbulence: A Long way from preheating to equilibrium}",
    eprint = "hep-ph/0210202",
    archivePrefix = "arXiv",
    doi = "10.1103/PhysRevLett.90.121301",
    journal = "Phys. Rev. Lett.",
    volume = "90",
    pages = "121301",
    year = "2003"
}

@article{Chesler:2008hg,
    author = "Chesler, Paul M. and Yaffe, Laurence G.",
    title = "{Horizon formation and far-from-equilibrium isotropization in supersymmetric Yang-Mills plasma}",
    eprint = "0812.2053",
    archivePrefix = "arXiv",
    primaryClass = "hep-th",
    doi = "10.1103/PhysRevLett.102.211601",
    journal = "Phys. Rev. Lett.",
    volume = "102",
    pages = "211601",
    year = "2009"
}

@article{Chesler:2010bi,
    author = "Chesler, Paul M. and Yaffe, Laurence G.",
    title = "{Holography and colliding gravitational shock waves in asymptotically AdS$_{5}$ spacetime}",
    eprint = "1011.3562",
    archivePrefix = "arXiv",
    primaryClass = "hep-th",
    doi = "10.1103/PhysRevLett.106.021601",
    journal = "Phys. Rev. Lett.",
    volume = "106",
    pages = "021601",
    year = "2011"
}

@article{Chesler:2009cy,
    author = "Chesler, Paul M. and Yaffe, Laurence G.",
    title = "{Boost invariant flow, black hole formation, and far-from-equilibrium dynamics in N = 4 supersymmetric Yang-Mills theory}",
    eprint = "0906.4426",
    archivePrefix = "arXiv",
    primaryClass = "hep-th",
    doi = "10.1103/PhysRevD.82.026006",
    journal = "Phys. Rev. D",
    volume = "82",
    pages = "026006",
    year = "2010"
}

@article{Heller:2012je,
    author = "Heller, Michal P. and Janik, Romuald A. and Witaszczyk, Przemyslaw",
    title = "{A numerical relativity approach to the initial value problem in asymptotically Anti-de Sitter spacetime for plasma thermalization - an ADM formulation}",
    eprint = "1203.0755",
    archivePrefix = "arXiv",
    primaryClass = "hep-th",
    doi = "10.1103/PhysRevD.85.126002",
    journal = "Phys. Rev. D",
    volume = "85",
    pages = "126002",
    year = "2012"
}

@article{Heller:2012km,
    author = "Heller, Michal P. and Mateos, David and van der Schee, Wilke and Trancanelli, Diego",
    title = "{Strong Coupling Isotropization of Non-Abelian Plasmas Simplified}",
    eprint = "1202.0981",
    archivePrefix = "arXiv",
    primaryClass = "hep-th",
    reportNumber = "ICCUB-12-009",
    doi = "10.1103/PhysRevLett.108.191601",
    journal = "Phys. Rev. Lett.",
    volume = "108",
    pages = "191601",
    year = "2012"
}

@article{vanderSchee:2012qj,
    author = "van der Schee, Wilke",
    title = "{Holographic thermalization with radial flow}",
    eprint = "1211.2218",
    archivePrefix = "arXiv",
    primaryClass = "hep-th",
    reportNumber = "ITP-UU-12-36",
    doi = "10.1103/PhysRevD.87.061901",
    journal = "Phys. Rev. D",
    volume = "87",
    number = "6",
    pages = "061901",
    year = "2013"
}

@article{Heller:2013oxa,
    author = "Heller, Michal P. and Mateos, David and van der Schee, Wilke and Triana, Miquel",
    title = "{Holographic isotropization linearized}",
    eprint = "1304.5172",
    archivePrefix = "arXiv",
    primaryClass = "hep-th",
    doi = "10.1007/JHEP09(2013)026",
    journal = "JHEP",
    volume = "09",
    pages = "026",
    year = "2013"
}

@article{Heinz:2001xi,
    author = "Heinz, Ulrich W. and Kolb, Peter F.",
    editor = "Karsch, F. and Satz, H.",
    title = "{Early thermalization at RHIC}",
    eprint = "hep-ph/0111075",
    archivePrefix = "arXiv",
    doi = "10.1016/S0375-9474(02)00714-5",
    journal = "Nucl. Phys. A",
    volume = "702",
    pages = "269--280",
    year = "2002"
}

@article{Chesler:2015lsa,
    author = "Chesler, Paul M. and van der Schee, Wilke",
    title = "{Early thermalization, hydrodynamics and energy loss in AdS/CFT}",
    eprint = "1501.04952",
    archivePrefix = "arXiv",
    primaryClass = "nucl-th",
    reportNumber = "MIT-CTP-4636",
    doi = "10.1142/S0218301315300118",
    journal = "Int. J. Mod. Phys. E",
    volume = "24",
    number = "10",
    pages = "1530011",
    year = "2015"
}

@article{Chesler:2015fpa,
    author = "Chesler, Paul M. and Kilbertus, Niki and van der Schee, Wilke",
    title = "{Universal hydrodynamic flow in holographic planar shock collisions}",
    eprint = "1507.02548",
    archivePrefix = "arXiv",
    primaryClass = "hep-th",
    doi = "10.1007/JHEP11(2015)135",
    journal = "JHEP",
    volume = "11",
    pages = "135",
    year = "2015"
}

@article{Chesler:2015bba,
    author = "Chesler, Paul M.",
    title = "{Colliding shock waves and hydrodynamics in small systems}",
    eprint = "1506.02209",
    archivePrefix = "arXiv",
    primaryClass = "hep-th",
    doi = "10.1103/PhysRevLett.115.241602",
    journal = "Phys. Rev. Lett.",
    volume = "115",
    number = "24",
    pages = "241602",
    year = "2015"
}

@article{Chesler:2016ceu,
    author = "Chesler, Paul M.",
    title = "{How big are the smallest drops of quark-gluon plasma?}",
    eprint = "1601.01583",
    archivePrefix = "arXiv",
    primaryClass = "hep-th",
    doi = "10.1007/JHEP03(2016)146",
    journal = "JHEP",
    volume = "03",
    pages = "146",
    year = "2016"
}

@article{Grozdanov:2016zjj,
    author = "Grozdanov, Sa\v{s}o and van der Schee, Wilke",
    title = "{Coupling Constant Corrections in a Holographic Model of Heavy Ion Collisions}",
    eprint = "1610.08976",
    archivePrefix = "arXiv",
    primaryClass = "hep-th",
    reportNumber = "MIT-CTP-4850",
    doi = "10.1103/PhysRevLett.119.011601",
    journal = "Phys. Rev. Lett.",
    volume = "119",
    number = "1",
    pages = "011601",
    year = "2017"
}

@article{Folkestad:2019lam,
    author = "Folkestad, Asmund and Grozdanov, Sa\v{s}o and Rajagopal, Krishna and van der Schee, Wilke",
    title = "{Coupling Constant Corrections in a Holographic Model of Heavy Ion Collisions with Nonzero Baryon Number Density}",
    eprint = "1907.13134",
    archivePrefix = "arXiv",
    primaryClass = "hep-th",
    reportNumber = "MIT-CTP/5136",
    doi = "10.1007/JHEP12(2019)093",
    journal = "JHEP",
    volume = "12",
    pages = "093",
    year = "2019"
}

@article{Heller:2016gbp,
    author = "Heller, Michal P.",
    title = "{Holography, Hydrodynamization and Heavy-Ion Collisions}",
    eprint = "1610.02023",
    archivePrefix = "arXiv",
    primaryClass = "hep-th",
    doi = "10.5506/APhysPolB.47.2581",
    journal = "Acta Phys. Polon. B",
    volume = "47",
    pages = "2581",
    year = "2016"
}

@inproceedings{Heinz:2002un,
    author = "Heinz, Ulrich W. and Kolb, Peter F.",
    title = "{Two RHIC puzzles: Early thermalization and the HBT problem}",
    booktitle = "{18th Winter Workshop on Nuclear Dynamics}",
    eprint = "hep-ph/0204061",
    archivePrefix = "arXiv",
    month = "4",
    year = "2002"
}

@article{Kolb:2003dz,
    author = "Kolb, Peter F. and Heinz, Ulrich W.",
    editor = "Hwa, Rudolph C. and Wang, Xin-Nian",
    title = "{Hydrodynamic description of ultrarelativistic heavy ion collisions}",
    eprint = "nucl-th/0305084",
    archivePrefix = "arXiv",
    reportNumber = "SUNY-NTG-03-06",
    pages = "634--714",
    month = "5",
    year = "2003"
}

@article{Heinz:2004pj,
    author = "Heinz, Ulrich W.",
    editor = "Bracco, Mirian and Chiapparini, Marcelo and Ferreira, Erasmo and Kodama, Takeshi",
    title = "{Thermalization at RHIC}",
    eprint = "nucl-th/0407067",
    archivePrefix = "arXiv",
    doi = "10.1063/1.1843595",
    journal = "AIP Conf. Proc.",
    volume = "739",
    number = "1",
    pages = "163--180",
    year = "2004"
}

@book{Casalderrey-Solana:2011dxg,
    author = "Casalderrey-Solana, Jorge and Liu, Hong and Mateos, David and Rajagopal, Krishna and Wiedemann, Urs Achim",
    title = "{Gauge/String Duality, Hot QCD and Heavy Ion Collisions}",
    eprint = "1101.0618",
    archivePrefix = "arXiv",
    primaryClass = "hep-th",
    reportNumber = "CERN-PH-TH-2010-316, MIT-CTP-4198, ICCUB-10-202",
    doi = "10.1017/9781009403504",
    isbn = "978-1-00-940350-4, 978-1-00-940349-8, 978-1-00-940352-8, 978-1-139-13674-7",
    publisher = "Cambridge University Press",
    year = "2014"
}

@article{Mueller:1999fp,
    author = "Mueller, Alfred H.",
    title = "{Toward equilibration in the early stages after a high-energy heavy ion collision}",
    eprint = "hep-ph/9906322",
    archivePrefix = "arXiv",
    reportNumber = "CU-TP-941",
    doi = "10.1016/S0550-3213(99)00502-7",
    journal = "Nucl. Phys. B",
    volume = "572",
    pages = "227--240",
    year = "2000"
}

@article{Mueller:2002gd,
    author = "Mueller, A. H. and Son, D. T.",
    title = "{On the Equivalence between the Boltzmann equation and classical field theory at large occupation numbers}",
    eprint = "hep-ph/0212198",
    archivePrefix = "arXiv",
    reportNumber = "CU-TP-1081, INT-PUB-02-56",
    doi = "10.1016/j.physletb.2003.12.047",
    journal = "Phys. Lett. B",
    volume = "582",
    pages = "279--287",
    year = "2004"
}

@article{Arnold:2003rq,
    author = "Arnold, Peter Brockway and Lenaghan, Jonathan and Moore, Guy D.",
    title = "{QCD plasma instabilities and bottom up thermalization}",
    eprint = "hep-ph/0307325",
    archivePrefix = "arXiv",
    doi = "10.1088/1126-6708/2003/08/002",
    journal = "JHEP",
    volume = "08",
    pages = "002",
    year = "2003"
}

@article{Kurkela:2014tea,
    author = "Kurkela, Aleksi and Lu, Egang",
    title = "{Approach to Equilibrium in Weakly Coupled Non-Abelian Plasmas}",
    eprint = "1405.6318",
    archivePrefix = "arXiv",
    primaryClass = "hep-ph",
    reportNumber = "CERN-PH-TH-2014-093",
    doi = "10.1103/PhysRevLett.113.182301",
    journal = "Phys. Rev. Lett.",
    volume = "113",
    number = "18",
    pages = "182301",
    year = "2014"
}

@article{Kurkela:2018vqr,
    author = {Kurkela, Aleksi and Mazeliauskas, Aleksas and Paquet, Jean-Fran\c{c}ois and Schlichting, S\"oren and Teaney, Derek},
    title = "{Effective kinetic description of event-by-event pre-equilibrium dynamics in high-energy heavy-ion collisions}",
    eprint = "1805.00961",
    archivePrefix = "arXiv",
    primaryClass = "hep-ph",
    doi = "10.1103/PhysRevC.99.034910",
    journal = "Phys. Rev. C",
    volume = "99",
    number = "3",
    pages = "034910",
    year = "2019"
}

@article{Blaizot:2011xf,
    author = "Blaizot, Jean-Paul and Gelis, Francois and Liao, Jin-Feng and McLerran, Larry and Venugopalan, Raju",
    title = "{Bose--Einstein Condensation and Thermalization of the Quark Gluon Plasma}",
    eprint = "1107.5296",
    archivePrefix = "arXiv",
    primaryClass = "hep-ph",
    doi = "10.1016/j.nuclphysa.2011.10.005",
    journal = "Nucl. Phys. A",
    volume = "873",
    pages = "68--80",
    year = "2012"
}

@article{Tanji:2017suk,
    author = "Tanji, Naoto and Venugopalan, Raju",
    title = "{Effective kinetic description of the expanding overoccupied Glasma}",
    eprint = "1703.01372",
    archivePrefix = "arXiv",
    primaryClass = "hep-ph",
    doi = "10.1103/PhysRevD.95.094009",
    journal = "Phys. Rev. D",
    volume = "95",
    number = "9",
    pages = "094009",
    year = "2017"
}

@article{Boguslavski:2023jvg,
    author = "Boguslavski, Kirill and Kurkela, Aleksi and Lappi, Tuomas and Lindenbauer, Florian and Peuron, Jarkko",
    title = "{Limiting attractors in heavy-ion collisions}",
    eprint = "2312.11252",
    archivePrefix = "arXiv",
    primaryClass = "hep-ph",
    doi = "10.1016/j.physletb.2024.138623",
    journal = "Phys. Lett. B",
    volume = "852",
    pages = "138623",
    year = "2024"
}

@article{Mazeliauskas:2018yef,
    author = {Mazeliauskas, Aleksas and Berges, J{\"u}rgen},
    title = "{Prescaling and far-from-equilibrium hydrodynamics in the quark-gluon plasma}",
    eprint = "1810.10554",
    archivePrefix = "arXiv",
    primaryClass = "hep-ph",
    doi = "10.1103/PhysRevLett.122.122301",
    journal = "Phys. Rev. Lett.",
    volume = "122",
    number = "12",
    pages = "122301",
    year = "2019"
}

@article{Strickland:2017kux,
    author = "Strickland, Michael and Noronha, Jorge and Denicol, Gabriel",
    title = "{Anisotropic nonequilibrium hydrodynamic attractor}",
    eprint = "1709.06644",
    archivePrefix = "arXiv",
    primaryClass = "nucl-th",
    doi = "10.1103/PhysRevD.97.036020",
    journal = "Phys. Rev. D",
    volume = "97",
    number = "3",
    pages = "036020",
    year = "2018"
}

\end{document}